\documentclass[letterpaper,twocolumn,10pt]{article}
\usepackage{usenix2019_v3}
\usepackage{comment}

\newcommand{\name}{\textsc{Themis}}
\newcommand{\agent}{\textsc{Agent}}
\newcommand{\arbiter}{\textsc{Arbiter}}

\usepackage{algorithm, algorithmicx, algpseudocode}
\usepackage[T1]{fontenc}
\usepackage{amsmath, amssymb, amsthm, amsfonts}
\usepackage{multirow}
\usepackage{hyperref}
\usepackage{color}
\usepackage{xcolor}
\usepackage{alltt}
\usepackage[draft]{pgf}
\usepackage{tikz}
\usetikzlibrary{backgrounds}
\usetikzlibrary{calc}
\usetikzlibrary{positioning}
\usetikzlibrary{automata}
\usetikzlibrary{shapes}
\usetikzlibrary{arrows}
\usetikzlibrary{shadows}
\usetikzlibrary{decorations.pathreplacing}
\usetikzlibrary{decorations.text}
\usetikzlibrary{shapes.multipart}
\usetikzlibrary{fit}
\usetikzlibrary{patterns}
\usepackage{subfig}

\usepackage[medium, compact]{titlesec}
\usepackage[font=footnotesize]{caption}
\titlespacing*{\section}{1pt}{3.5pt}{2pt}
\titlespacing*{\subsection}{1pt}{3pt}{1.5pt}
\titlespacing*{\subsubsection}{1pt}{3pt}{1.5pt}

\newcommand{\mytilde}{\raise.17ex\hbox{$\scriptstyle\mathtt{\sim}$}}

\newcommand*\squared[1]{\tikz[baseline=(char.base)]{%
            \node[shape=rectangle,fill=black,inner sep=1pt] (char) {\color{white}#1};}}

\newenvironment{compactlist}{
	\begin{list}{{$\bullet$}}{
			\setlength\partopsep{0pt}
			\setlength\parskip{0pt}
			\setlength\parsep{0pt}
			\setlength\topsep{2pt}
			\setlength\itemsep{4pt}
			\setlength{\itemindent}{\leftmargin}
			\setlength{\leftmargin}{0pt}
		}
	}{
\end{list}
}

\newcommand{\myvec}[1]{\protect\overrightarrow{#1}}

\def\ie{{i.e.}}

\newenvironment{proofof}[1]{\vspace{0.1in}\noindent{\sc Proof of #1.}}{\hfill\qed}

\begin{document}

\title{\Large \bf {\name}: Fair and Efficient GPU Cluster Scheduling}
\author{
Kshiteej Mahajan\\
University of Wisconsin - Madison
\and
Arjun Balasubramanian\\
University of Wisconsin - Madison
\and
Arjun Singhvi\\
University of Wisconsin - Madison
\and
Shivaram Venkataraman\\
University of Wisconsin - Madison
\and
Aditya Akella\\
University of Wisconsin - Madison
\and
Amar Phanishayee\\
Microsoft Research
\and
Shuchi Chawla\\
University of Wisconsin - Madison
}

\maketitle

\noindent{\bf Abstract:}
Modern distributed machine learning (ML) training workloads benefit
significantly from leveraging GPUs. However, significant contention
ensues when multiple such workloads are run atop a shared cluster of
GPUs. A key question is how to fairly apportion GPUs across
workloads. We find that established cluster scheduling disciplines are
a poor fit because of ML workloads' unique attributes: ML jobs have
long-running tasks that need to be gang-scheduled, and their
performance is sensitive to tasks' relative placement. %

We propose \name{}, a new scheduling framework for ML training
workloads. It's GPU allocation policy enforces that ML workloads
complete in {\em a finish-time fair} manner, a new notion we
introduce. To capture placement sensitivity and ensure efficiency,
\name{} uses a two-level scheduling architecture where ML workloads
bid on available resources that are offered in an {\em auction} run by
a central arbiter. Our auction design allocates GPUs to winning bids
by trading off efficiency for fairness in the short term, but ensuring
finish-time fairness in the long term. Our evaluation on a production trace
shows that \name{} can improve fairness by more than $2.25X$ and is ~$5\%$ to $250$\% more cluster efficient in comparison to state-of-the-art
schedulers.

\section{Introduction}

With the widespread success of machine learning (ML) for tasks
such as object detection, speech recognition, and machine translation,
a number of enterprises are now incorporating ML models into their
products.  Training individual ML models is time- and
resource-intensive with each training job typically executing in
parallel on a number of GPUs.

With different groups in the same organization training ML models, it
is beneficial to consolidate GPU resources into a shared
cluster. Similar to existing clusters used for large scale data
analytics, shared GPU clusters for ML have a number of operational
advantages, e.g., reduced development overheads, lower costs for
maintaining GPUs, etc. However, today, there are no ML
workload-specific mechanisms to share a GPU cluster in a {\em fair}
manner.

Our conversations with cluster operators indicate that fairness is
crucial; specifically, that sharing an ML cluster becomes
attractive to users only if they have the appropriate \emph{sharing
  incentive}. That is, if there are a
total $N$ users sharing a cluster $C$, every user's performance should
be no worse than using a private cluster of size $\frac{C}{N}$. Absent
such incentive, users are either forced to sacrifice performance and
suffer long wait times for getting their ML jobs scheduled, or
abandon shared clusters and deploy their own expensive hardware.

Providing sharing incentive through fair scheduling mechanisms has
been widely studied in prior cluster scheduling frameworks, e.g.,
Quincy~\cite{quincy}, DRF~\cite{drf}, and
Carbyne~\cite{carbyne}. However, these techniques were designed for
big data workloads, and while they are used widely to manage GPU
clusters today, they are far from effective.

The key reason is that ML workloads have unique characteristics that
make existing ``fair'' allocation schemes actually {\em unfair}. 
First, unlike batch analytics
workloads, ML jobs have {\em long running tasks} that need to be
scheduled together, i.e., gang-scheduled. Second, each task in a job
often runs for a number of iterations while synchronizing model
updates at the end of each iteration. This frequent communication
means that jobs are {\em placement-sensitive}, i.e., placing all the
tasks for a job on the same machine or the same rack can lead to
significant speedups. Equally importantly, as we show, ML jobs differ
in their relative placement-sensitivity
(Section~\ref{sec:placement_pref}).

In Section~\ref{sec:fair}, we show that having long-running tasks means that
established schemes such as DRF -- which aims to equally allocate the
GPUs released upon task completions -- can arbitrarily violate sharing
incentive. We show that even if GPU resources were
released/reallocated on fine time-scales~\cite{tiresias}, placement
sensitivity means that jobs with same aggregate resources could have
widely different performance, violating sharing incentive. Finally,
heterogeneity in placement sensitivity means that existing scheduling
schemes {\em also violate} Pareto efficiency and envy-freedom, two
other properties that are central to fairness~\cite{varian}.

Our scheduler, \name{}, address these challenges, and supports sharing
incentive, Pareto efficiency, and envy-freedom for ML workloads. It
multiplexes a GPU cluster across \emph{ML applications}
(Section~\ref{sec:motivation}), or apps for short, where every app
consists of one or more related ML jobs, each running with different
hyper-parameters, to train an accurate model for a given task. To
capture the effect of long running tasks and placement sensitivity,
\name{} uses a new long-term fairness metric, \emph{finish-time
  fairness}, which is the ratio of the running time in a shared cluster
with $N$ apps to running alone in a $\frac{1}{N}$ cluster. \name{}'s
goal is thus to minimize the maximum finish time fairness across all
ML apps while efficiently utilizing cluster GPUs. We achieve this
goal using two key ideas.

First, we propose to widen the API between ML apps and the scheduler
to allow apps to specify placement preferences. We do this by
introducing the notion of a round-by-round auction.
\name{} uses leases to account for long-running ML tasks, and auction
rounds start when leases expire.  At the start of a round, our
scheduler requests apps for their finish-time fairness metrics, and
makes all available GPUs visible to  a fraction of apps
that are currently farthest in terms of their fairness metric.  Each
such app has the opportunity to {\em bid} for subsets of these GPUs as
a part of an auction; bid values reflect the app's new (placement
sensitive) finish time fairness metric from acquiring different GPU
subsets. A central arbiter determines the global winning bids to
maximize the aggregate improvement in  the finish time fair metrics across all bidding
apps. Using auctions means that we need to ensure that apps are
truthful when they bid for GPUs. Thus we use a \emph{partial
  allocation} auction that incentivizes truth telling, and ensures
Pareto-efficiency and envy-freeness by design.

While a far-from-fair app may lose an auction round, perhaps because it
placed less ideally than another app, its bid values for subsequent
auctions naturally increase (because a losing app's finish time
fairness worsens), thereby improving the odds of it winning future rounds. Thus, our
approach converges to fair allocations over the long term, while
staying efficient and placement-sensitive in the short term.

Second, we present a two-level scheduling design that contains a
centralized inter-app scheduler at the bottom level, and a narrow API
to integrate with existing hyper-parameter tuning frameworks at the top
level. A number of existing frameworks such as
Hyperdrive~\cite{hyperdrive} and HyperOpt~\cite{hyperopt} can
intelligently apportion GPU resources between various jobs in a single
app, and in some cases also terminate a job early if its progress is
not promising. Our design allows apps to directly use such existing
hyper parameter tuning frameworks. We describe how \name{}
accommodates various hyper-parameter tuning systems and how its API is
exercised in extracting relevant inputs from apps when running
auctions.

We implement \name{} atop Apache YARN 3.2.0, %
and evaluate by replaying workloads from a large
enterprise trace. Our results show
that \name{} is atleast $2.25X$ better than state-of-the-art schedulers while also improving cluster efficiency by ~$5$\% to $250$\%.
To further understand our scheduling decisions, we perform an event-driven simulation using the same
trace, and our results show that \name{} offers greater benefits when we increase the fraction of network intensive apps, 
and the cluster contention.

\section{Motivation} \label{sec:motivation}

We start by defining the terminology used in
the rest of the paper. 
We then study the unique properties of ML workload 
traces from a ML training GPU cluster 
at a large
software 
company. We end by stating our goals based on our trace
analysis and conversations with the cluster operators.

\subsection{Preliminaries} \label{sec:motivation_defs}
We define an ML app, or simply an ``app'', as a collection of one
or more ML model {\em training} jobs. Each app corresponds to a user training an
ML model for a high-level goal, such as speech recognition or 
object detection. Users train these models knowing the appropriate hyper-parameters 
(in which case there is just a single job in the app), 
or they train a closely related set of models ($n$ jobs) that explore hyper-parameters 
such as learning rate, momentum etc.~\cite{hyperdrive,hyperband} 
to identify and train the best target model for the activity at hand.

Each job's constituent work is performed by a number of parallel
\emph{tasks}.  At any given time, all of a job's tasks collectively
process a {\em mini-batch} of training data; we assume that the size of
the batch is fixed for the duration of a job. Each task typically
processes a subset of the batch, and, starting from an initial version
of the model, executes multiple iterations of the underlying learning
algorithm to improve the model. We assume all jobs use the
popular synchronous SGD~\cite{google-sync-async}.

We consider the finish time of an app to be when the best model and
relevant hyper-parameters have been identified. Along the course of
identifying such a model, the app may decide to terminate some of its
constituent jobs early~\cite{hyperdrive,hyperopt}; such jobs may be
exploring hyper-parameters that are clearly sub-optimal (the jobs'
validation accuracy improvement over iterations is significantly worse
than other jobs in the same app). For apps that contain a single job,
finish time is the time taken to train this model to a target
accuracy or maximum number of iterations.

\subsection{Characterizing Production ML Apps} \label{sec:empirical_study}

\begin{figure*}[!t]
	\begin{minipage}[t]{0.24\textwidth}
	  \includegraphics[width=\textwidth]{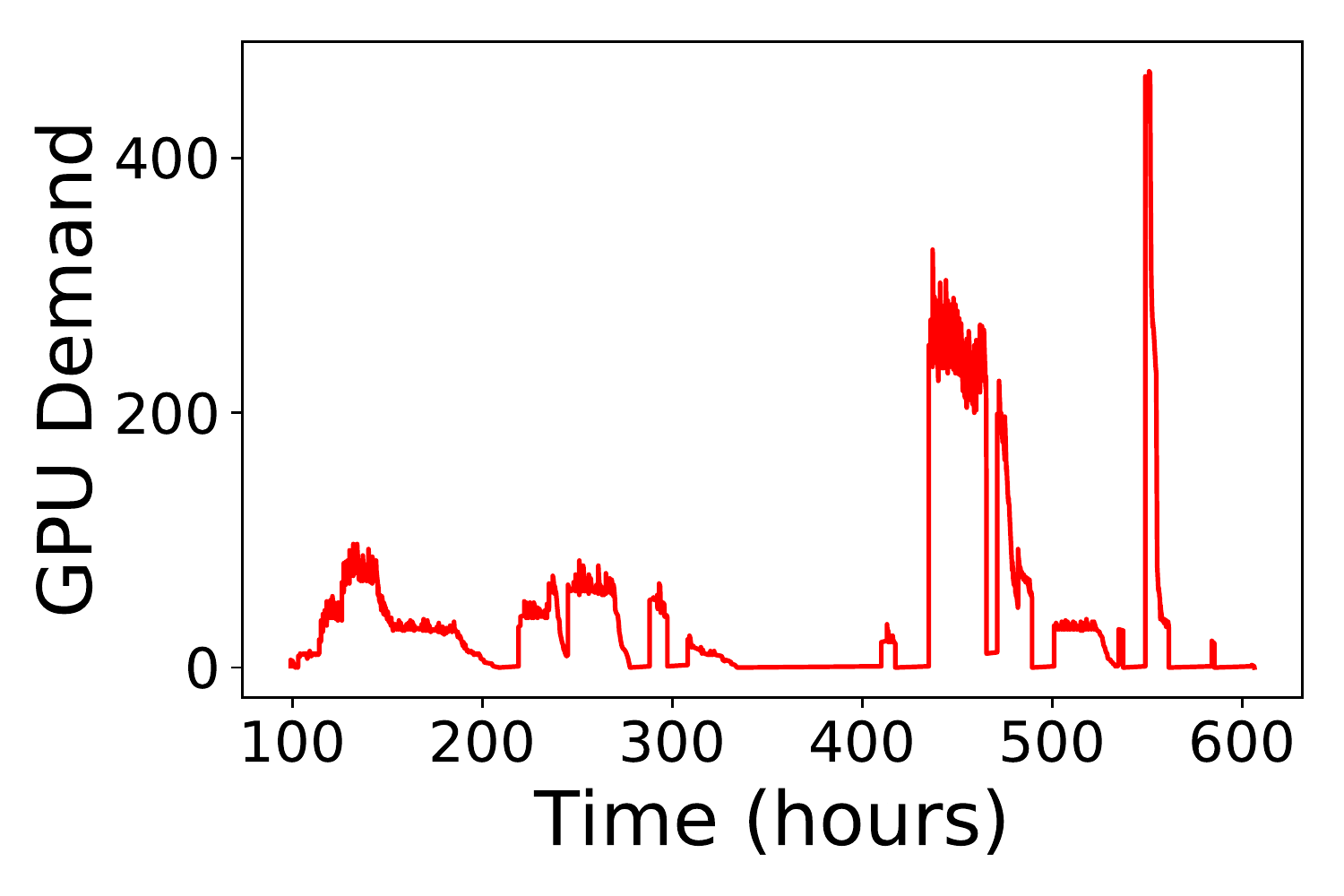} 
  	\caption{Aggregate GPU demand of ML apps over time}
	  \label{fig:gpu-contention-in-cluster}
   \end{minipage}
	\begin{minipage}[t]{0.24\textwidth}
		\includegraphics[width=\textwidth]{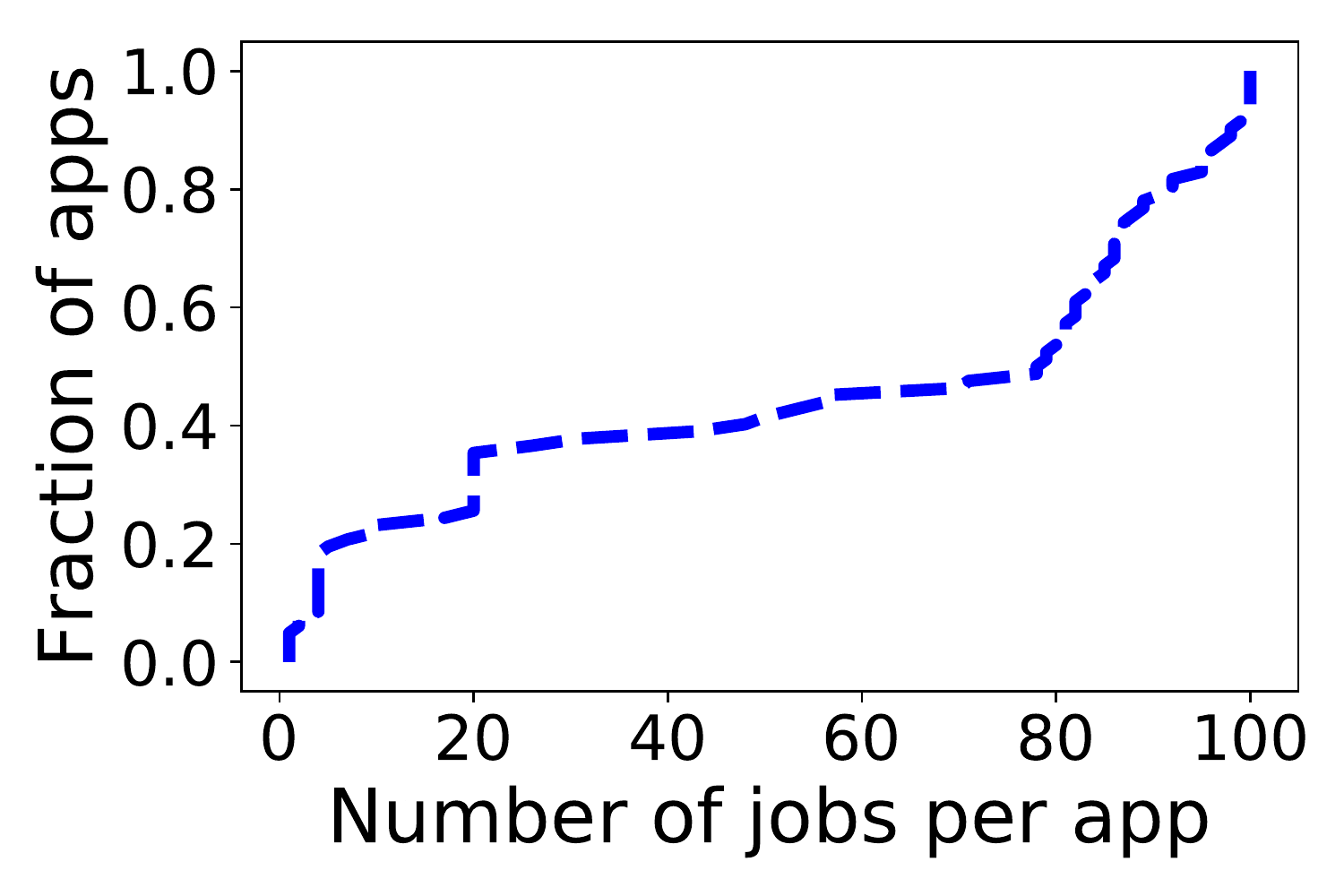}
		\caption{Job count distribution across different apps}
		\label{fig:jobs-per-app}
	\end{minipage}%
  \hspace{0.1cm}
	\begin{minipage}[t]{0.24\textwidth}
		\includegraphics[width=\textwidth]{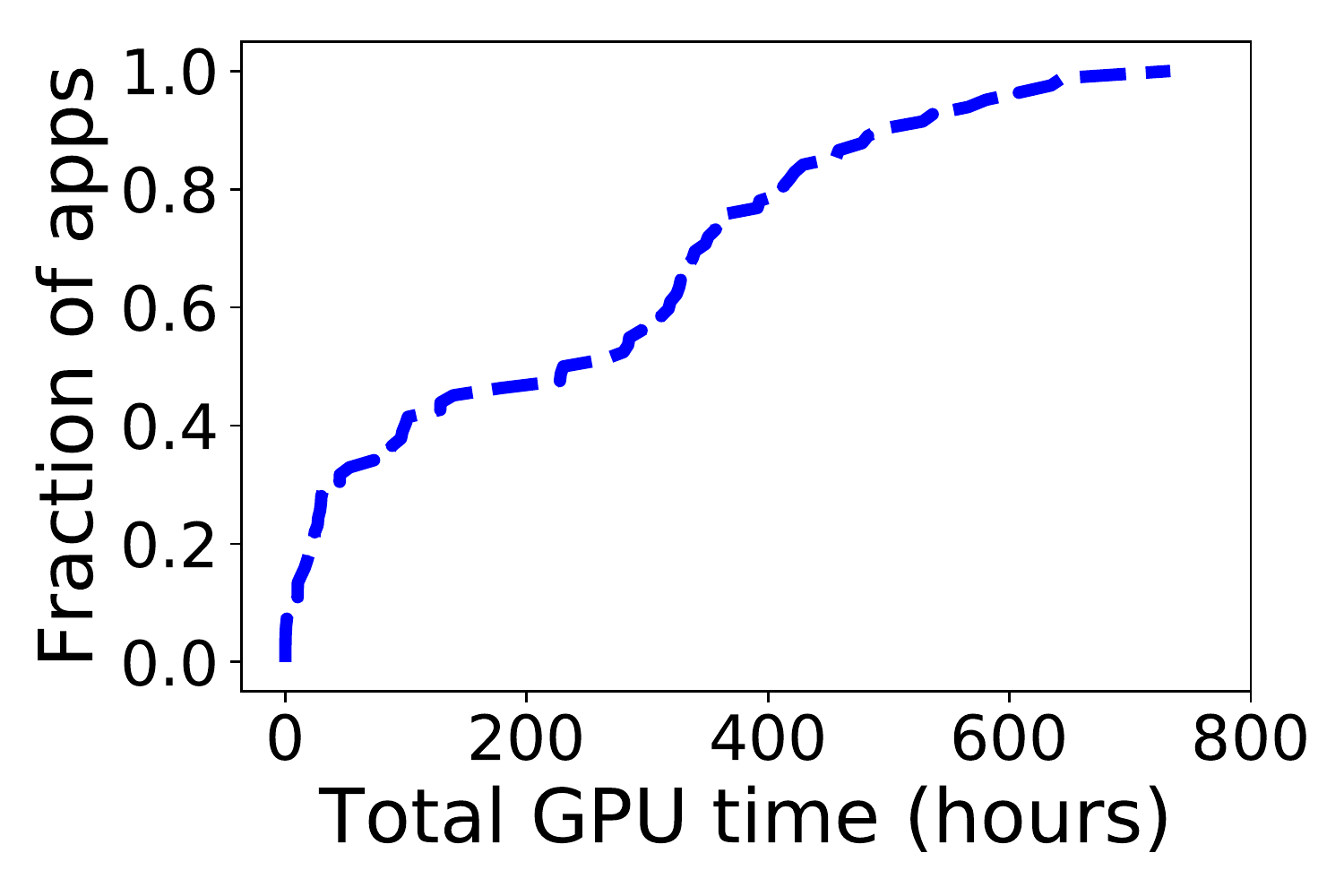}
		\caption{ML app time ( = total GPU time across all jobs in app) distribution}
		\label{fig:cdf-app-gpu-time}
	\end{minipage}
  \hspace{0.1cm}
	\begin{minipage}[t]{0.24\textwidth}
		\includegraphics[width=\textwidth]{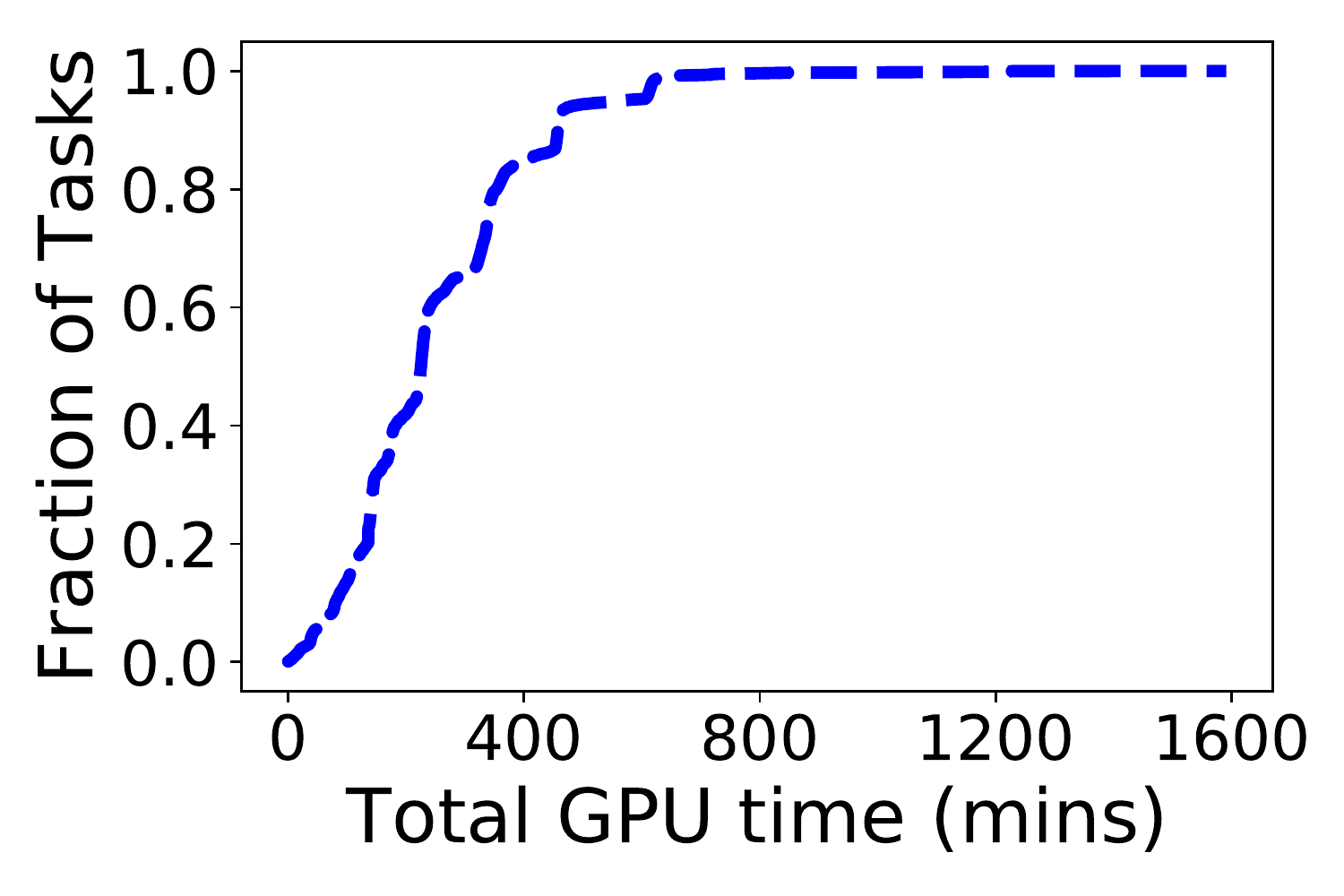}
		\caption{Distribution of Task GPU times}
		\label{fig:job-runtime-across-apps}
	\end{minipage}%
  \vspace{-0.25in}
\end{figure*}

We perform an analysis of the properties of GPU-based ML training
workloads by analyzing workload traces obtained from a large internet
company.  The GPU cluster we study
supports over 5000 unique users.  We restrict our analysis to a subset of the trace
that contains 85 ML training apps submitted using a hyper-parameter tuning framework.

 GPU clusters are known to be heavily
contented~\cite{philly_trace_analysis}, and we find this also holds
true in the subset of the trace of ML apps we consider
(Figure~\ref{fig:gpu-contention-in-cluster}). 
For instance, we see that GPU demand 
is bursty and the average GPU demand is \textasciitilde 50 GPUs. %

We also use the trace to provide a first-of-a-kind view into the
characteristics of ML apps.  As mentioned in Section
\ref{sec:motivation_defs}, apps may either train a single model to
reach a target accuracy (1 job) or may use the cluster to explore
various hyper-parameters for a given model (n jobs).
Figure~\ref{fig:jobs-per-app} shows that \textasciitilde10\% of the
apps have 1 job, and around \textasciitilde90\% of the apps perform
hyper-parameter exploration with as many as 100 jobs (median of 75
jobs).  Interestingly, there is also a significant variation in the
number of hyper-parameters explored ranging from a few tens to about a
hundred (not shown).

We also measure the {\em GPU time} of all ML apps in the trace.  
If an app uses 2 jobs with 2 GPUs each for a period of 10 minutes, 
then the task GPU times would be 10 minutes each, 
the job GPU times would be 20 minutes each, and the app GPU time would be 40 GPU minutes.
Figure~\ref{fig:cdf-app-gpu-time} and
Figure~\ref{fig:job-runtime-across-apps} show the long running nature
of ML apps: the median app takes 11.5 GPU days and the median task
takes 3.75 GPU hours. There is a wide diversity with a significant
fraction of jobs and apps that are more than 10X shorter and many that
are more than 10X longer.

From our analysis we see that ML apps are heterogeneous in terms of
resource usage, and number of jobs submitted. Running times are also
heterogeneous, but at the same time much longer than, e.g., running
times of big data analytics jobs (typical a few
hours~\cite{graphene}). Handling such heterogeneity can be challenging
for scheduling frameworks, and the long running nature may make
controlling app performance particularly difficult in a shared setting
with high contention.

We next discuss how some of these challenges manifest in practice from
both cluster user and cluster operator perspectives, and how that
leads to our design goals for \name{}.

\subsection{Our Goal}

Our many conversations
with operators of GPU clusters revealed a common sentiment, reflected in the following quote:

{\small\emph{``
    We were scheduling with a balanced approach ... with guidance to 
    'play nice'. Without firm guard rails, however, there were always individuals who would ignore the
    rules and dominate the capacity.
    ''}}

\rightline{{\small --- An operator  of a large GPU cluster}}

With long app durations, users who dominate capacity impose high waiting times on many other users. Some such users are forced to ``quit'' the cluster as reflected in this quote:

{\small\emph{``Even with existing fair sharing schemes, we do find
    users frustrated with the inability to get their work done in a
    timely way... The frustration frequently reaches the point
    where groups attempt or succeed at buying their own hardware
    tailored to their needs.  ''}}

\rightline{{\small --- An operator  of a large GPU cluster}}

While it is important to design a cluster scheduler that ensures
efficient use of highly contended GPU resources, the above indicates
that it is perhaps equally, if not more important for the scheduler to
allocate GPU resources in a fair manner across many diverse ML apps;
in other words, roughly speaking, the scheduler's goal should be to
allow all apps to execute their work in a ``timely way''.

In what follows, we explain using examples, measurements, and analysis, why existing
fair sharing approaches when applied to ML clusters fall short of the
above goal, which we formalize next. We identify the need both for a
new  fairness metric, and for a new scheduler
architecture and API that supports resource division according to the
metric.

\section{Finish-Time Fair Allocation}
\label{sec:fair}

We present additional unique attributes of ML apps and discuss how
they, and the above attributes, affect existing fair sharing schemes.

\subsection{Fair Sharing Concerns for ML Apps} \label{sec:fair_sharing_concerns}
The central question is - given $R$ GPUs in a cluster $C$ and $N$
ML apps, what is a {\em fair way} to divide the GPUs. %

As mentioned above, cluster operators indicate that the primary
concern for users sharing an ML cluster is performance isolation that
results in ``timely completion''. We formalize this as: if $N$ ML Apps
are sharing a cluster then an app should not run slower on the shared
cluster compared to a dedicated cluster with $\frac{1}{N}$ of the
resources. Similar to prior work~\cite{drf}, we refer to this property
as {\em sharing incentive} (SI). Ensuring sharing incentive for ML
apps is our primary design goal.

In addition, resource allocation mechanisms must satisfy two other
basic properties that are central to fairness~\cite{varian}: Pareto
Efficiency (PE) and Envy-Freeness (EF)~\footnote{Informally, a Pareto
  Efficient allocation is one where no app's allocation can be
  improved without hurting some other app.  And, envy-freeness means
  that no app should prefer the resource allocation of an other app.}

While prior systems like Quincy~\cite{quincy}, DRF~\cite{drf} etc. aim
at providing SI, PE and EF, we find that they are ineffective for ML
clusters as they fail to consider {\em the long durations of ML tasks}
and {\em placement preferences of ML apps}.

\subsubsection{ML Task Durations}
We empirically study task durations in ML apps and show how they
affect the applicability of existing fair sharing schemes.

Figure~\ref{fig:job-runtime-across-apps} shows the distribution of
task durations for ML apps in a production cluster.
We note that the tasks are, in general,
very long, with the median task roughly 3.75 hours long. This is in
stark contrast with, e.g., big data analytics jobs, where tasks are
typically much shorter in duration~\cite{sparrow}.

State of the art fair allocation schemes such as DRF~\cite{drf}
provide instantaneous resource fairness. Whenever resources become
available, they are allocated to the task from an app with the least
current share. For big data analytics, where task durations are short,
this approximates instantaneous resource fairness, as frequent task
completions serve as opportunities to redistribute resources. However,
blindly applying such schemes to ML apps can be disastrous: running
the much longer-duration ML tasks to completion could lead to newly
arriving jobs waiting inordinately long for resources. This leads to
violation of SI for late-arriving jobs.

Recent ``attained-service'' based schemes address this problem with
DRF. In~\cite{tiresias}, for example, GPUs are leased for a certain
duration, and when leases expire, available GPUs are 
given to the  
job that received the least GPU time thus far; this is the ``least
attained service'', or LAS allocation policy. While this scheme avoids
the starvation problem above for late-arriving jobs, it still violates
all key fairness properties because it is placement-unaware, an issue
we discuss next.

\subsubsection{Placement Preferences}
\label{sec:placement_pref}
Next, we empirically study placement preferences of ML
apps.  We use examples to show how ignoring these preferences in
fair sharing schemes violates key properties of fairness. %

\begin{figure}[!t]
	\centering
	\includegraphics[width=0.6\columnwidth]{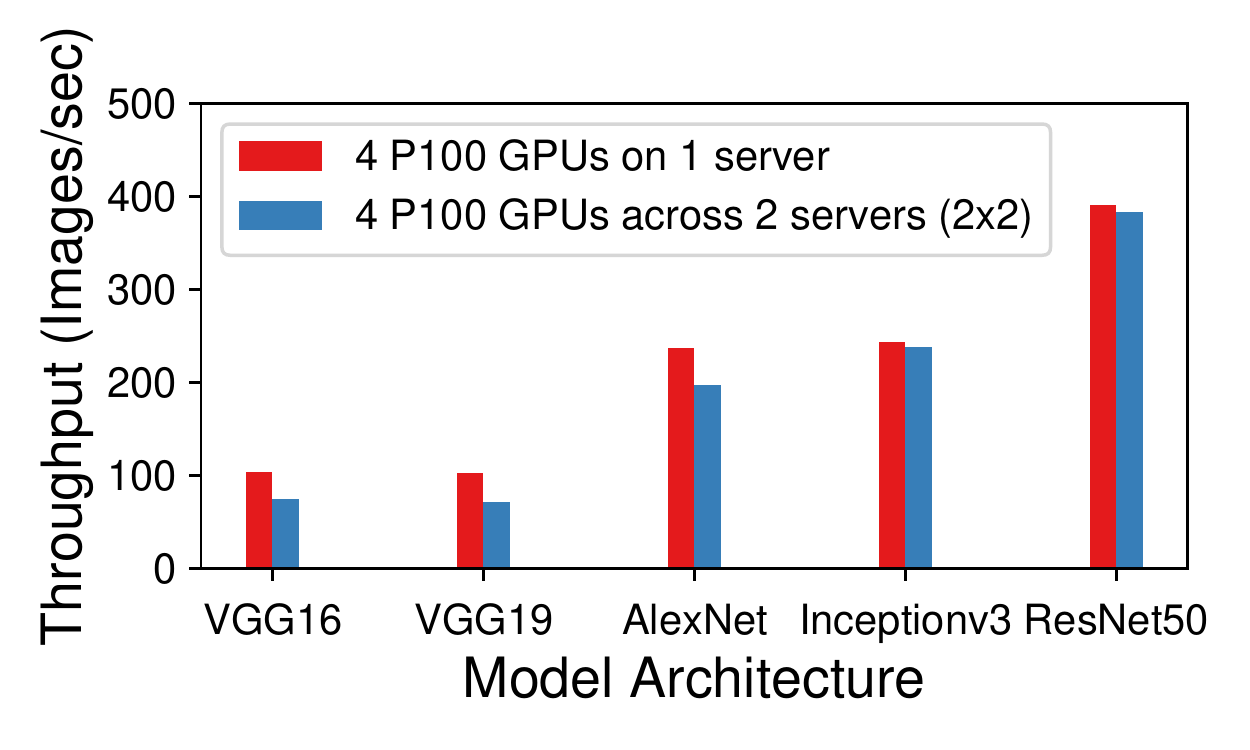} 
	\caption{Effect of GPU resource allocation on job throughput.}
	\label{fig:placement}
\end{figure}

{\noindent \bf Many apps, many preference patterns:} ML cluster users
today train a variety of ML apps across domains like computer vision,
NLP and speech recognition.  These models have significantly different
model architectures, and more importantly, different placement
preferences arising from different computation, communication needs.
For example, as shown in Figure~\ref{fig:placement}, VGG16 has a
strict machine-local task placement preference while Resnet50 does
not.  This preference inherently stems from the fact that VGG-like
architectures have very large number of parameters and incur greater
overheads for updating gradients over the network.

We use examples to show the effect of placement on DRF's allocation
strategy. Similar examples and conclusions apply for the LAS allocation
scheme.

\begin{figure}[!t]
	\centering
	\includegraphics[width=\columnwidth]{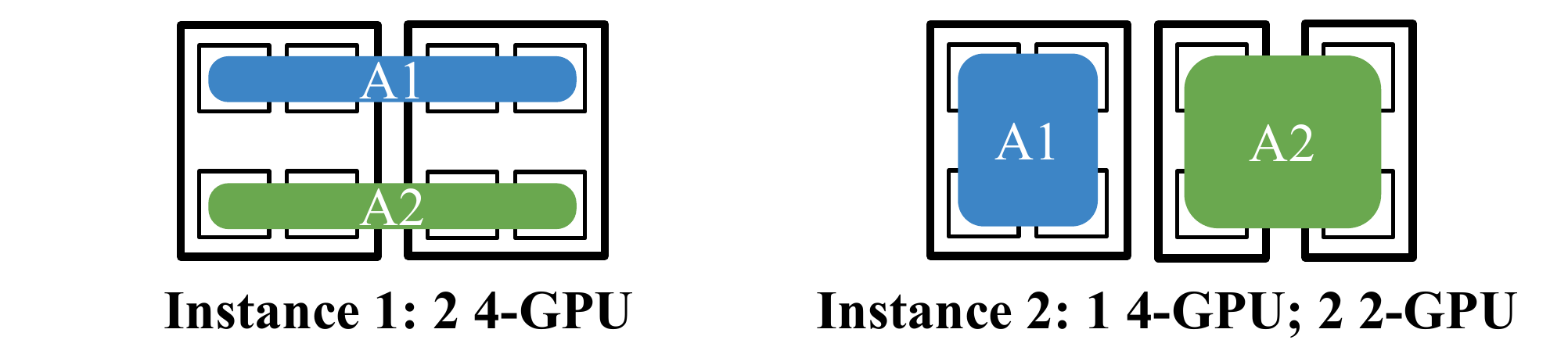} 
	\caption{By ignoring placement preference, DRF violates sharing incentive.}
	\label{fig:drf-violates}
\end{figure}

{\noindent \bf Ignoring placement affects SI: example:} Consider the
Instance 1 in Figure~\ref{fig:drf-violates}.  In this example, there are
two placement sensitive ML apps - $A_{1}$ reflecting VGG16, and
$A_{2}$ reflecting VGG19.  Each ML app has just one job in it with 
$4$ tasks and the cluster has two $4$ GPU machines. 
As shown above, given the same number of GPUs both apps prefer GPUs to
be in the same server than spread across servers.

For this example, DRF~\cite{drf} equalizes the dominant resource share
of both the apps under resource constraints and allocates $4$ GPUs to
each ML app.  In Instance 1 of Figure~\ref{fig:drf-violates} we show
an example of a valid DRF allocation.  Both apps get the same type of
placement with GPUs spread across servers. This allocation violates SI
for both apps as their performance would be better if each app just
had its own dedicated server.

{\noindent \bf Ignoring placement affects PE, EF: example:} Consider
Instance 2 in Figure~\ref{fig:drf-violates} with two apps -
$A_{1}$ (Inceptionv3) which is not placement sensitive and
$A_{2}$ (VGG16) which is placement sensitive. Each
app has one job with four tasks and and the cluster has two machines: one $4$ GPU 
and two $2$ GPU.

Now consider the allocation in Instance 2, where $A_{1}$ is allocated on the $4$
GPU machine  whereas $A_{2}$ is allocated across the 2 GPU machines. %
This allocation violates EF, because $A_{2}$ would prefer $A_{1}$'s
allocation. It also violates PE because swapping the two apps'
allocation would improve $A_{2}$'s performance without hurting
$A_{1}$.

In fact, we can formally show that:

\noindent{\textbf{Theorem 3.1.}}
    Existing fair schemes (DRF, LAS) ignore
    placement preferences and violate SI, PE, EF for ML apps.

\noindent{\textit{Proof}} Refer to Appendix.

In summary, existing schemes fail to provide fair sharing guarantees
as they are unaware of ML app characteristics. Instantaneous fair
schemes such as DRF fail to account for long task durations. While
least-attained service schemes overcome that limitation, neither
approach's input encodes placement preferences.
Correspondingly, the fairness metrics used -
i.e., dominant resource share (DRF) or attained service (LAS) - do not
capture placement preferences.

This motivates the need for a new placement-aware fairness metric, and
corresponding scheduling discipline. Our
observations about ML task durations imply that, like LAS, our fair
allocation discipline should not depend on rapid task
completions, but instead should operate over longer time scales.

\subsection{Metric: Finish-Time Fairness}
We propose a new metric called as finish-time fairness, $\rho$.
{$\rho = \frac{T_{sh}}{T_{id}}$}. 

$T_{id}$ is the {\em independent finish-time} and 
$T_{sh}$ is the {\em shared finish-time}. 
$T_{sh}$ is the finish-time of the app 
in the shared cluster and it encompasses the slowdown due to the 
placement and any queuing delays that an app experiences in getting scheduled in the shared cluster. The worse the placement the higher is the value of $T_{sh}$. 
$T_{id}$, is the finish-time of the 
ML app in its own independent and exclusive $\frac{1}{N}$ share of 
the cluster.
Given the above definition, sharing incentive for an ML app can be attained if $\rho \le 1$.

To ensure this, it is necessary for the allocation mechanism to estimate the values of
$\rho$ for different GPU allocations. Given the difficulty in predicting how various apps will react
to different allocations, it is intractable for the scheduling engine to predict or determine the values of
$\rho$.

Thus we propose a new wider interface between the app and
the allocation engine that can allow the app to express a {\em preference} for each allocation. 
We propose that apps can encode this information as a table. In Table~\ref{tab:example-bid-vector},
each column has a permutation of a potential 
GPU allocation and the estimate of $\rho$ on receiving this allocation. 
We next describe how the scheduling engine can use this to provide fair allocations.

\begin{table}
  \begin{scriptsize}
	\begin{center}
  \begin{tabular}{ c|c|c|c } 
    $\myvec{G}$ & $[0,0]$ & $[0,1] = [1,0]$ & $[1,1]$ \\ 
    \hline
    $\rho$ & $\rho_{old}$ & $\frac{200}{400} = \frac{1}{2}$ & $\frac{100}{400} = \frac{1}{4}$ \\
	\end{tabular}
	\end{center}
  \vspace{-0.1in}
	\caption{Example table of bids sent from apps to the scheduler}
 	\label{tab:example-bid-vector}
  \vspace{-0.05in}
\end{scriptsize}
\end{table}

\subsection{Mechanism: Partial Allocation Auctions}
\floatname{algorithm}{Pseudocode}
\begin{algorithm}[t!]

\begin{scriptsize}
\begin{algorithmic}[1]

\State Applications \{$\textit{A}_i$\} \Comment{set of apps}
\State Bids \{$\mathbb{\rho}_i(.)$\} \Comment{valuation function for each app $i$}
\State Resources $\myvec{R}$ \Comment{resource set available for auction}
\State Resource Allocations \{$\myvec{G}_i$\} \Comment{resource allocation for each app $i$}

\Statex

\Procedure{auction}{\{$\textit{A}_i$\}, \{$\mathbb{\rho}_i(.)$\}, $\myvec{R}$} 
    \State $\myvec{G}_{i,pf}$ = arg max $\prod_i 1/\mathbb{\rho}_i(\myvec{G_{i}})$  \Comment{proportional fair (pf) allocation per app $i$}
    \State $\myvec{G}^{-i}_{j,pf}$ = arg max $\prod_{j != i} 1/\mathbb{\rho}_j(\myvec{G_{j}})$  \Comment{pf allocation per app $j$ without app $i$}
    \State $c_i$ = $\frac{\prod_{j != i} 1/\mathbb{\rho}_j(\myvec{G}_{j,pf})}{\prod_{j != i} 1/\mathbb{\rho}_j(\myvec{G}^{-i}_{j,pf})}$
    \State $\myvec{G_i}$ = $c_i$ * $\myvec{G}_{i,pf}$ \Comment{allocation per app $i$} 
    \State $\myvec{L}$ = $\sum_i$ $1 - c_i$ * $\myvec{G}_{i,pf}$  \Comment{aggregate leftover resource}
    \State return \{$\myvec{G_i}$\}, $\myvec{L}$
\EndProcedure

\Procedure{roundByRoundAuctions}{\{$\textit{A}_i$\}, \{$\mathbb{\rho}_i(.)$\}}
    \While{True}
    \State \textsc{onResourceAvailableEvent} $\vec{R'}$:
    \State \{$A_{i}^{sort}$\} = \textsc{sort}(\{${A_i}$\}) on $\rho_{i}^{current}$
    \State \{$A_{i}^{filter}$\} = get top $1 - f$ fraction of apps from \{$A_{sort}$\}
    \State \{$\rho_{i}^{filter}(.)$\} = get updated $\rho(.)$ from apps in \{$A_{i}^{filter}$\}
    \State \{$\myvec{G_i^{filter}}$\}, $\myvec{L}$ = \textsc{auction}(\{$A_{i}^{filter}$\}, \{$\rho_{i}^{filter}(.)$\}, $\vec{R'}$)
    \State \{$A_{i}^{unfilter}$\} = $\{A_{i}\} - \{A_{i}^{filter}\}$
    \State allocate $\myvec{L}$ to \{$A_{i}^{unfilter}$\} at random
    \EndWhile
\EndProcedure

\end{algorithmic}
\end{scriptsize}
\caption{Finish-Time Fair Policy}
\label{alg:pam}
\end{algorithm}

The finish-time fairness $\rho_{i}(.)$ for an ML app $A_{i}$ is a function  
of the GPU allocation $\vec{G_{i}}$ that it receives. 
The allocation policy takes these $\rho_{i}(.)$'s as inputs and 
outputs allocations $\vec{G_{i}}$. 

Intuitively the scheduling engine would pick the apps that can most
benefit from an allocation and assign resources to that app. In other
words allocations which can decrease an app's $\rho_{i}$ will be
preferred.  However, one of the main challenges in a two level design
where apps submit their respective $\rho_{i}$'s is that apps could
potentially return {\em false information} to boost the chances that
they receive a particular allocation. Our conversations with cluster operators indicate that apps request for more resources than required and the require monitoring (``\emph{We also monitor the usage. If they don’t use it, we reclaim it and pass it on to the next approved project}'').
Thus a simple scheme like sorting the apps based on their reported $\rho$ values would fail to provide
another key property, namely, {\em strategy proofness} (SP).

To address this challenge, we propose to use {\em auctions} in \name{}.
We begin by describing a simple mechanism that runs a single-round auction and then extend to a round-by-round
mechanism that also considers online updates.

\subsubsection{One-Shot Auction}
Details of the inputs necessary to run the auction are given first, 
followed by how the auction works given these inputs. 

{\noindent \bf Inputs: Resources and Bids.} $\vec{R}$ represents 
the total GPU resources to be auctioned, where each element is $1$ and the number of dimensions is the number of GPUs to be auctioned.

Each ML app bids for these resources. 
The bid for each ML app consists of the estimated 
finish-time fair metric ($\rho_{i}$) values for several 
different GPU allocations ($\vec{G}_{i}$). 
Each element in $\vec{G}_{i}$ can be $\{0, 1\}$.  
A set bit implies that GPU is allocated to the app. Example of a bid can be seen in Table~\ref{tab:example-bid-vector}.

{\noindent \bf Auction Overview. } To ensure that the auction can
provide strategy proofness, we propose using a {\em partial
  allocation} auction (PA) mechanism~\cite{nopaymentfair}.  Partial
allocation auctions have been shown to incentivize truth telling and
are an appropriate fit 
for modeling subsets of indivisible goods to be auctioned across apps. 
Pseudocode~\ref{alg:pam}, line
$5$ shows the PA mechanism.  There are two aspects to auctions that
are described next.

{\noindent \bf 1. Initial allocation.} PA starts by calculating an
intrinsically proportionally-fair allocation $\vec{G_{i, pf}}$ for
each app $A_{i}$ by maximizing the product of the valuation functions
\ie~the finish-time fair metric values for all apps
(Pseudocode~\ref{alg:pam}, line $6$). Such an allocation ensures that
it is not possible to increase the allocation of an app without
decreasing the allocation of at least another app (satisfying PE~\cite{nopaymentfair}).

{\noindent \bf 2. Incentivizing Truth Telling.} To induce truthful
reporting of the bids, the PA mechanism allocates app $A_{i}$ only a
fraction $c_{i} < 1$ of $A_{i}$'s proportional fair allocation
$\vec{G_{i, pf}}$, and takes $1 - c_{i}$ as a {\em hidden payment}
(Pseudocode~\ref{alg:pam}, line $9$). The $c_i$ is directly
proportional to the decrease in collective valuation of the other
bidding apps in a market with and without app $A_{i}$
(Pseudocode~\ref{alg:pam}, line $8$). This yields the final allocation
$\vec{G_{i}}$ for app $A_{i}$ (Pseudocode~\ref{alg:pam}, line $9$).

Note that the final result, $\vec{G_{i}}$ is not a market-clearing allocation and there could be 
unallocated GPUs $\vec{L}$ that are leftover from hidden payments. 
Hence, PA is not work-conserving. Thus, while the one-shot auction provides a number of properties
related to fair sharing it does not ensure SI is met.

  \noindent{\textbf{Theorem 3.2.}}
    The one-shot partial allocation auction guarantees SP, PE and EF,
    but does not provide SI.

\noindent{\textit{Proof}} Refer to Appendix.

The intuitive reason for this is that the partial allocation mechanism does not
explicitly ensure that $\rho \le 1$ across apps.  To address this we
next look at multi-round auctions that can maximize sharing incentive
for ML apps.

\subsubsection{Multi-round auctions}
To maximize sharing incentive and to ensure work conservation, our goal is to ensure $\rho \le 1$ for as many apps as possible. 
We do this using three key ideas described below. 

{\noindent \bf Round-by-Round Auctions: } 
With round-by-round auctions, the outcome of an allocation from an 
auction is binding only for a {\em lease} duration. 
At the end of this lease, the freed GPUs are re-auctioned. 
This also handles the online case as any auction is triggered 
on a {\em resource available event}. This takes care of 
app failures and arrivals, as well as cluster reconfigurations. 

At the beginning of each round of auction, the policy solicits 
updated valuation functions $\rho(.)$ from the apps. 
The estimated work and the placement preferences for the case of 
ML apps are typically time varying. 
This also makes our policy adaptive to such changes. 

{\noindent \bf Round-by-Round Filtering: } To maximize the number of
apps with $\rho \le 1$, at the beginning of each round of auctions we
filter the $1 - f$ fraction of total active apps with the 
greatest values of current estimate of their finish-time
fair metric $\rho$. Here, $f \in (0,1)$ is a system-wide parameter.

This has the effect of restricting the auctions to the apps that are
at risk of not meeting SI. Also, this restricts the
auction to a smaller set of apps which reduces contention for
resources and hence results in smaller hidden payments. It also makes
the auction computationally tractable.

Over the course of many rounds, filtering maximizes the number of apps
that have SI. Consider a far-from-fair app $i$ that lost an auction
round. It will appear in future rounds with much greater likelihood
relative to another less far-from-fair app $k$ that won the auction
round. This is because, the winning app $k$ was allocated resources;
as a result, it will see its $\rho$ improve over time; thus, it will
eventually not appear in the fraction $1-f$ of not-so-fairly-treated
apps that participate in future rounds. In contrast, $i$'s $\rho$
will increase due to the waiting time, and thus it will continue to appear in future
rounds. Further an app that loses multiple rounds will eventually lose
its lease on all resources and make no further progress, causing its
$\rho$ to become unbounded. The next auction round the app
participates in will likely see the app's bid winning, because any
non-zero GPU allocation to that app will lead to a huge improvement in
the app's valuation.

As $f \rightarrow 1$, our policy provides greater guarantee on
SI. However, this increase in SI comes at the cost of 
efficiency. This is because $f \rightarrow 1$ restricts the set of
apps to which available GPUs will be allocated; with $f \rightarrow 0$
available GPUs can be allocated to apps that benefit most from better placement,
which improves efficiency at the risk of violating SI.

{\noindent \bf Leftover Allocation: } At the end of each round
we have leftover GPUs due to hidden payments.  We allocate
these GPUs at random to the apps that did
not participate in the auction in this round. Thus our overall
scheme is work-conserving.

Overall, we prove that:

  \noindent{\textbf{Theorem 3.3.}}
  Round-by-round auctions preserve the PE, EF and SP
  properties of partial auctions and maximize SI.

  {\noindent \textit Proof.}
    Refer to Appendix.

To summarize, in \name{} we propose a new finish-time fairness metric that
captures fairness for long-running, placement sensitive ML apps. To perform
allocations, we propose using a multi-round partial allocation auction that incentivizes
truth telling and provides pareto efficient, envy free allocations. By filtering
the apps considered in the auction, we maximize sharing incentive and hence satisfy
all the properties necessary for fair sharing among ML applications.

\section{System Design} \label{sec:system_design}

We first list design requirements for an ML cluster scheduler taking into account the fairness
metric and auction mechanism described in Section~\ref{sec:fair}, and the
implications for the \name{} scheduler architecture. Then, we
discuss the {\em API} between the scheduler and the hyper-parameter optimizers.

\subsection{Design Requirements}

{\noindent \bf Separation of visibility and allocation of resources.}
Core to our partial allocation mechanism is the abstraction of making
available resources visible to a number of apps but allocating
each resource exclusively to a single app. As we argue below, existing
scheduling architectures couple these concerns and thus necessitate
the design of a new scheduler. %

{\noindent \bf Integration with hyper-parameter tuning systems.}  Hyper-parameter optimization systems such as Hyperband~\cite{hyperband}, Hyperdrive~\cite{hyperdrive} have their own schedulers that decide the resource allocation and execution schedule for the jobs within those apps. We refer to these as app-schedulers. One of our goals in \name{} is to integrate with these systems with minimal modifications to app-schedulers.

These two requirements guide our design of a new {\em two-level semi-optimistic} scheduler
 and a set of corresponding abstractions to support hyper-parameter tuning systems.

\subsection{\name{} Scheduler Architecture}
Existing scheduler architectures are either pessimistic or fully optimistic and both these
approaches are not suitable for realizing multi-round auctions. We first describe their shortcomings
and then describe our proposed architecture.

\subsubsection{Need for a new scheduling architecture}
Two-level pessimistic schedulers like Mesos~\cite{mesos} enforce
pessimistic concurrency control. This means that visibility and
allocation go hand-in-hand at the granularity of a single app. There
is restricted single-app visibility as available resources are
partitioned by a mechanism internal to the lower-level (i.e., cross-app) 
scheduler and offered only to a single app at a time.  The
tight coupling of visibility and allocation makes it infeasible to
realize round-by-round auctions where resources need to be visible to
many apps but allocated to just one app.

Shared-state fully optimistic schedulers like Omega~\cite{omega} enforce fully optimistic concurrency control.
This means that visibility and allocation go hand-in-hand at the granularity of multiple apps. There
is full multi-app visibility as all cluster resources and their state is made visible to all apps.
Also, all apps contend for resources and resource allocation decisions are made by multiple apps at
the same time using transactions. %
This coupling of visibility and allocation in a makes it hard to
realize a global policy like finish-time fairness and also leads to
expensive conflict resolution (needed when multiple apps contend for
the same resource) when the cluster is highly contented, which is
typically the case in shared GPU clusters.

Thus, the properties required by multi-round auctions, i.e., multi-app resource visibility and single-app resource allocation granularity, makes existing architectures
ineffective. %

\subsubsection{Two-Level Semi-Optimistic Scheduling} %

\begin{figure}[t]
    \centering \includegraphics[width=\columnwidth]{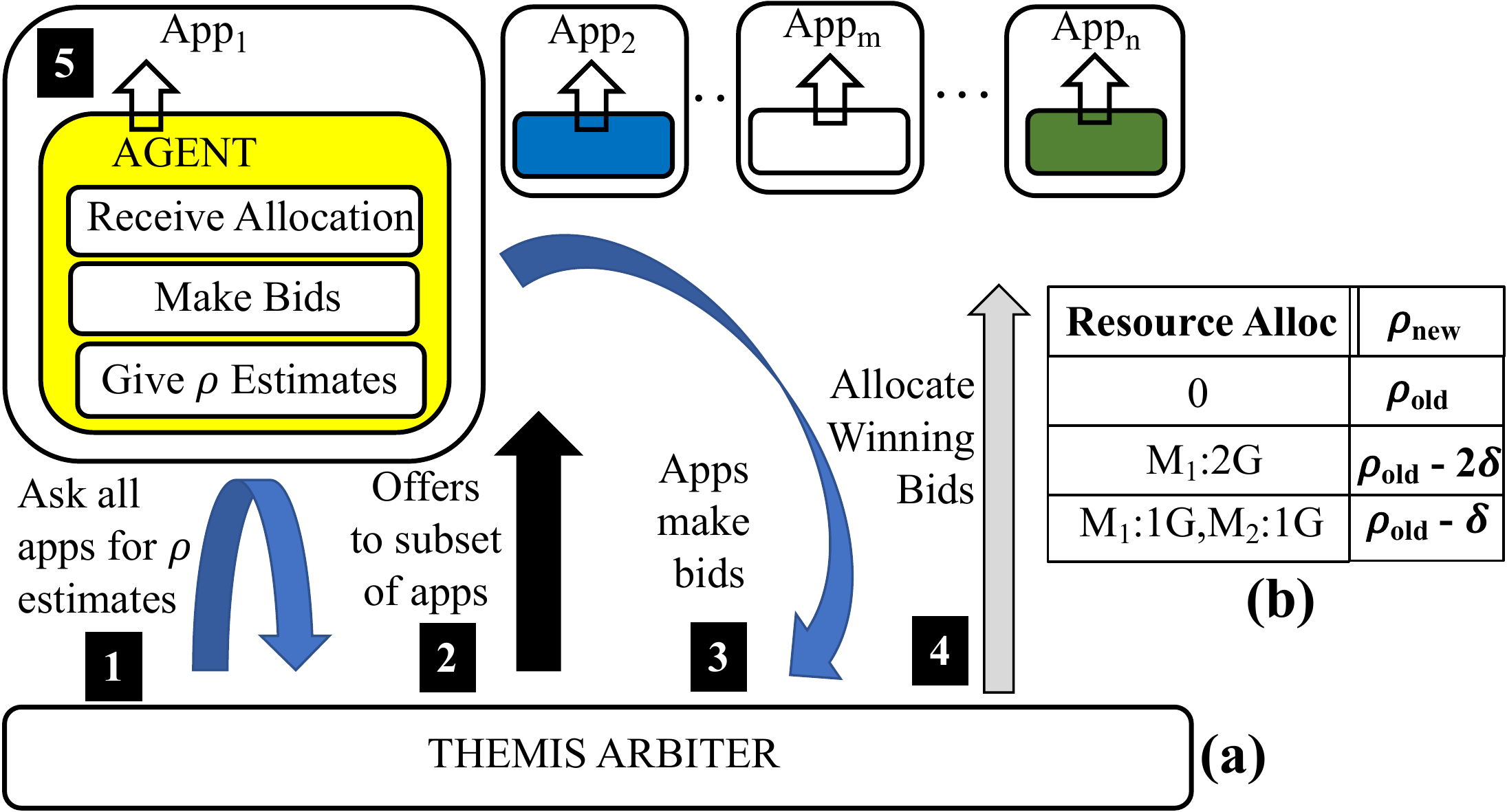}
    \caption{
        \name{} Design. (a) Sequence of events in \name{} - starts with a resource available event and ends with resource allocations. 
        (b) Shows a typical bid valuation table an App submits to \arbiter{}. 
        Each row has a subset of the complete resource allocation and the improved value of $\rho_{new}$. 
    }
    \label{fig:overview}
    \vspace{-0.1cm}
\end{figure}

The two-levels in our scheduling architecture comprise of multiple app-schedulers and a cross-app
scheduler that we call the \textsc{arbiter}. The \arbiter{} has our scheduling logic.
The top level per-app schedulers are minimally modified
to interact with the \arbiter.
Figure~\ref{fig:overview} shows our architecture.

Each GPU in a \name{}-managed cluster has a lease associated with it.
The lease decides the duration of ownership of the GPU for an app.
When a lease expires, the resource is made available for
allocation. \name{}'s \arbiter{} pools available resources and runs a round of the auctions
described earlier.
During each such round, the resource allocation proceeds in 5 steps spanning 2 phases
(shown in Figure~\ref{fig:overview}):

The first phase, called the {\em visibility phase}, spans steps 1--3.

\squared{1} The \arbiter{} asks all apps for current finish-time fair metric estimates.
\squared{2} The \arbiter{} initiates auctions, and makes the same
non-binding resource-offer of the available resources to 
to a 
fraction $f \in [0,1]$ of ML apps with worst finish-time fair metrics (according to round-by-round filtering described earlier).
To minimize changes in the ML app scheduler to participate in
auctions, \name{} introduces an \agent{} that is co-located with each
ML app scheduler. The \agent{} serves as an intermediary between the
ML app and the \arbiter{}. %
\squared{3} The apps examine the resource offer in parallel. Each app's \agent{} then replies with a single
         bid that contains preferences for desired resource allocations. 

The second phase, {\em allocation phase}, spans steps 4--5.
\squared{4} The \arbiter{}, upon receiving all the bids for this round,  picks winning
bids according to previously described partial allocation algorithm and leftover allocation scheme. It then notifies each \agent{} of its winning
allocation (if any). 
\squared{5} The \agent{} propagates the allocation to the ML app
scheduler, which can then decide the allocation among
constituent jobs.

In sum, the two phase resource allocation means that our scheduler
enforces {\em semi-optimistic concurrency control}.
Similar to 
fully optimistic concurrency control, there is
multi-app visibility as the cross-app scheduler offers 
resources to multiple apps concurrently. 
At the same time, similar
to pessimistic concurrency control, the resource allocations are
conflict-free guaranteeing exclusive access of a resource to every
app.

 To enable preparation of bids in step \squared{3}, \name{} implements
        a narrow API from the ML app scheduler to the \agent{} that
        enables propagation of app-specific information. An \agent{}'s bid
        contains a {\em valuation function} ($\rho(.)$) that
        provides, for each resource subset, an
        estimate of the finish-time fair metric the app
        will achieve with the allocation of the resource subset.
        We describe how this is calculated next.

\subsection{\agent{} and AppScheduler Interaction}
\label{sec:api}
An \agent{} co-resides with an app 
to aid participation in auctions. 
We now describe how \agent{}s prepare bids based on inputs provided by apps, 
the API between an \agent{} and its app, and how \agent{}s integrate with
current hyper-parameter optimization schedulers.

\subsubsection{Single-Job ML Apps}
For ease of explanation, we first start with the simple case of an ML
app that has just one ML training job which can use at most
$job\_demand_{max}$ GPUs.  We first look at calculation of the
finish-time fair metric, $\rho$.  We then look at a multi-job app
example so as to better understand the various steps and interfaces in
our system involved in a multi-round auction.

{\noindent \bf Calculating $\rho(\myvec{G})$.} 
Equation~\ref{eq:rho-job} shows the steps for calculating $\rho$ 
for a single job given a GPU allocation of $\myvec{G}$ 
in a cluster $C$ with $R_{C}$ GPUs. When calculating $\rho$ 
we assume that the allocation $\myvec{G}$ is binding till job completion.
\vspace{-0.15in}
\begin{equation} \label{eq:rho-job}
    \begin{split}
        \rho(\myvec{G}) &= T_{sh}(\myvec{G})/T_{id} \\
        T_{sh} &= T_{current} - T_{start} + \\
        & iter\_left * iter\_time(\myvec{G}) \\
        T_{id} &= T_{cluster} * N_{avg} \\ 
        iter\_time(\myvec{G}) &= \frac{iter\_time\_serial * \mathcal{S}(\myvec{G})}{min(||\myvec{G}||_{1}, job\_demand_{max})} \\ 
        T_{cluster} &= \frac{iter\_total * iter\_serial\_time}{min(R_C, job\_demand_{max})}
    \end{split}
\end{equation}

$T_{sh}$ is the shared finish-time and is a function of the allocation $\myvec{G}$ 
that the job receives. 
For the single job case, it has two terms. 
First, is the time elapsed ($= T_{current} - T_{start}$). Time elapsed also 
captures any queuing delays or starvation time. 
Second, is the time to execute remaining iterations which is the product of the number of iterations 
left ($iter\_left$) and the iteration time ($iter\_time(\myvec{G})$). 
$iter\_time(\myvec{G})$ depends on the allocation received. 
Here, we consider the common-case of the ML training job executing synchronous SGD. 
In synchronous SGD, the work in an iteration can be parallelized across multiple workers. 
Assuming linear speedup, this means that the iteration time is  
the serial iteration time ($iter\_time\_serial$) reduced by a factor of 
the number of GPUs in the allocation, $||\myvec{G}||_{1}$ or $job\_demand_{max}$ 
whichever is lesser. 
However, the linear speedup assumption is not true in the common case as network 
overheads are involved. We capture this via a slowdown penalty, $\mathcal{S}(\myvec{G})$, which 
depends on the placement of the GPUs in the allocation. Values for
$\mathcal{S}(\myvec{G})$ can typically be obtained by profiling the job offline for a few iterations.
The slowdown is captured as a multiplicative factor, $\mathcal{S}(\myvec{G}) \ge 1$, by which $T_{sh}$ is increased.

$T_{id}$ is the estimated finish-time in an independent $\frac{1}{N_{avg}}$ cluster. 
$N_{avg}$ is the average contention in the cluster and is the weighed average of 
the number of other apps in the system during the lifetime of the app. 
We approximate this as the finish-time of the app in the whole cluster, $T_{cluster}$ 
divided by the average contention. $T_{cluster}$ assumes linear speedup when 
the app executes with all the cluster resources $R_C$ or maximum app demand 
whichever is lesser. It also assumes no slowdown. Thus, it is approximated as $\frac{iter\_total * iter\_serial\_time}{min(R_C, job\_demand_{max})}$. 

\subsubsection{Generalizing to Mutiple-Job ML Apps}
ML app schedulers for hyper-parameter optimization systems typically go from 
aggressive exploration of hyper-parameters to aggressive exploitation of best hyper-parameters. 
While there are a number of different algorithms for choosing the best hyper-parameters~\cite{hyperband,hyperopt} to run, we focus on early stopping criteria as this affects the
finish time of ML apps.

As described in prior work~\cite{vizier}, automatic stopping algorithms can be divided into two
categories: Successive Halving and Performance Curve Stopping. We next discuss how to compute $T_{sh}$ for each case.

{\noindent \bf Successive Halving} refers to schemes which start with a total time or iteration budget $B$ and
apportion that budget by periodically stopping jobs that are not promising. For example,
if we start with $n$ hyper parameter options, then each one is submitted as a job with a demand of $1$ GPU for a
fixed number of iterations $I$. After $I$ iterations, only the best $\frac{n}{2}$ ML training jobs
are retained and assigned a  maximum demand of $2$ GPUs for the same number of iterations $I$.
This continues until we are left with $1$ job with a maximum demand of $n$ GPUs. 
Thus there are a total of $log_{2}n$ {\em phases} in Successive Halving. This scheme is used in
Hyperband~\cite{hyperband} and Google Vizier~\cite{vizier}. 

We next describe how to compute the $T_{sh}$ and $T_{id}$ for successive halving.  
We assume that the given allocation $\myvec{G}$ lasts till app completion and the total time can be
computed by adding up the time the app spends for each phase. Consider the case of phase $i$ which has $J =
\frac{n}{2^{i-1}}$ jobs. Equation~\ref{eq:rho-halving} shows the calculation of $T_{sh(i)}$, the shared finish time of the phase.  
We assume a separation of concerns where the hyper-parameter optimizer can determine the optimal
allocation of GPUs {\em within a phase} and thus estimate the value of $\mathcal{S}(\myvec{G_{j}})$.
Along with $iter\_left$, $serial\_iter\_time$, the \agent{} can now estimate $T_{sh(j)}$ for each job in the phase.
We mark the phase as finished when the slowest or last job in the app finishes the phase ($max_{j}$). 
Then the shared finish time for the app is the sum of the finish times of all constituent phases.

To estimate the ideal finish-time we compute the total time to execute the app on the full cluster.
We estimate this using the budget $B$ which represents the aggregate work to be done and, as before, we
assume linear speedup to the maximum number of GPUs the app can use $app\_demand_{max}$.
\vspace{-0.11in}
\begin{equation} \label{eq:rho-halving}
    \begin{split}
        T_{sh(i)} &= max_{j} \{T(\myvec{G_{j}})\} \\
        T_{sh} &= \sum_{i} T_{sh(i)} \\
        T_{cluster} &= \frac{B}{min(R_C, app\_demand_{max})}\\
        T_{id} &= T_{cluster} * N_{avg}
    \end{split}
\end{equation}

The \agent{} generates $\rho$ using the above procedure for all possible subsets of $\{\myvec{G}\}$
and produced a bid table similar to the one shown in Table~\ref{tab:example-bid-vector} before.
The API between the \agent{} and hyper-parameter optimizer is shown in Figure~\ref{fig:api-example}
and captures the functions that need to implemented by the hyper-parameter optimizer.

{\noindent \bf Performance Curve Stopping} refers to schemes where the convergence curve of a job is
extrapolated to determine which jobs are more promising. This scheme is used by
Hyperdrive~\cite{hyperdrive} and Google Vizier~\cite{vizier}. Computing $T_{sh}$ proceeds
by calculating the finish time for each job that is
currently running by estimating the iteration at which the job will be terminated (Thus $T_{sh}$ is determined by the job that finishes last). As before, we assume that the given allocation $\myvec{G}$ lasts till app completion. Since the
estimations are usually probabilistic (i.e. the iterations at which the job will converge has an
error bar), we over-estimate and use the most optimistic convergence curve that results in the
maximum forecasted completion time for that job. As the job progresses, the estimates of the convergence
curve get more accurate and improving the estimated finish time $T_{sh}$. The API implemented by the
hyper-parameter optimizer is simpler and only involves getting a list of running jobs as shown in
Figure~\ref{fig:api-example}.

We next present an end-to-end example of a multi-job app showing our mechanism in action.

\subsubsection{End-to-end Example.} 
We now run through a simple example that exercises the various aspects of our API and the interfaces involved. 

Consider a $16$ GPU cluster and an ML app that has $4$ ML jobs and uses successive halving, running
along with 3 other ML apps in the same cluster. Each job in the app is tuning a different hyper-parameter and
the serial time taken per iteration for the jobs are $80,100,100,120$ seconds respectively.\footnote{The time per
iteration depends on the nature of the hyper-parameter being tuned. Some hyper-parameters like batch
size or quantization used affect the iteration time while others like learning rate don't.}
The total budget for the app is to $10,000$ seconds of GPU time and we assume the
$job\_demand_{max}$ is $8$ GPUs and $\mathcal{S}(\myvec{G}) = 1$. %

Given we start with $4$ ML jobs, the hyper-parameter optimizer divides this into 3 phases each having
$4, 2, 1$ jobs, resply. To evenly divide the budget across the phases, the hyper-parameter
optimizer budgets $\approx8,16,36$ iterations in each phase.
First we calculate the $T_{id}$ by considering the budget, total cluster size, and cluster
contention as: $\frac{10000 \times 4}{16} = 2500$s. 

Next we consider the computation of $T_{sh}$ assuming that $16$ GPUs are offered by the
\textsc{arbiter}. The \textsc{agent} now computes the bid for each subset of GPUs offered. Consider
the case with $2$ GPUs. In this case in the first phase we have $4$ jobs which are
serialized to run $2$ at a time. This leads to $T_{sh(1)} = (120 \times 8) + (80 \times 8) = 1600$
seconds. (Assume two 100s jobs run serially on one GPU, and the 80 and 120s jobs run serially on the
other. $T_{sh}$ is the time when the last job finishes.)

When we consider the next stage the hyper-parameter optimizer currently
does not know which jobs will be chosen for termination. We use the
\emph{median} job (in terms of per-iteration time) to estimate
$T_{sh(i)}$ for future phases. Thus, in the second phase we have $2$
jobs so we run one job on each GPU each of which we assume to take the
median $100$ seconds per iteration leading to
$T_{sh(2)} = (100 \times 16) = 1600$ seconds. Finally for the last
phase we have $1$ job that uses $2$ GPUs and runs for $36$ iterations
leading to $T_{sh(3)} = \frac{(100 \times 36)}{2} = 1800$ (again, the
``median'' jobs takes 100s per iteration). Thus
$T_{sh} = 1600+1600+1800 = 5000$ seconds, making $\rho =
\frac{5000}{2500} = 2$. 
Note that since placement did not matter here we considered any $2$ GPUs being used. Similarly
ignoring placement, the bids for the other allocations are
shown in Table~\ref{tab:bid-example}.

We highlight a few more points about our example above. If the jobs
that are chosen for the next phase do not match the median iteration
time then the estimates are revised in the next round of the
auction. For example if the jobs that are chosen for the next round
have iteration time $120,100$ then the above bid will be updated with
$T_{sh(2)} = (120 \times 16) = 3200$\footnote{Because the two jobs run on one
GPU each, and the 120s-per-iteration job is the last to finish in the
phase} and $T_{sh(3)} = \frac{(120 \times 36)}{2} = 2160$.
Further we also see that the $job\_demand_{max} = 8$ means that the $\rho$ value for $16$
GPUs does not linearly decrease from that of $8$ GPUs.

\begin{figure}[!t]
  \begin{minipage}[t]{\columnwidth}
    \begin{center}
    \begin{small}
    \begin{alltt}
class JobInfo(int itersRemaining,
              float avgTimePerIter,
              float localitySensitivity);
// Successive Halving
List<JobInfo> getJobsInPhase(int phase,
                             List<Int> gpuAlloc);
int getNumPhases();
// Performance Curve
List<JobInfo> getJobsRemaining(List<Int> gpuAlloc);
    \end{alltt}
    \end{small}
    \end{center}
  \end{minipage}
  \vspace{-0.1in}
  \caption{API between \agent{} and hyperparameter optimizer}
  \label{fig:api-example}
  \vspace{-0.05in}
\end{figure}

\begin{table}
\begin{scriptsize}
\begin{center} 
\begin{tabular}{ c|c|c|c|c|c|c } 
  $||\myvec{G}||_{1}$ & $0$ & $1$ & $2$ & $4$ & $8$ & $16$ \\ 
  \hline
  $\rho$ &     $\rho_{old}$ & $4$ & $2$ & $1$ & $0.5$ & $0.34$ \\ 
\end{tabular}
\end{center}
  \vspace{-0.1in}
 \caption{Example of bids submitted by \textsc{Agent}}
  \label{tab:bid-example}
  \vspace{-0.05in}
\end{scriptsize}
\end{table}

\section{Implementation}

We implement \name{} on top of a recent release of Apache Hadoop
YARN~\cite{yarn} (version $3.2.0$) which includes,
Submarine~\cite{submarine}, a new framework for running
 ML training jobs atop YARN. 
We modify the Submarine client to support submitting a group of ML
training jobs as required by hyper-parameter exploration apps.  
Once an app is submitted, it is
managed by a Submarine Application Master (AM) 
and we make changes to the Submarine AM to 
implement the ML
app scheduler (we use Hyperband~\cite{hyperband}) and
our \agent{}.

To prepare accurate bids, we implement a profiler in the AM that
parses TensorFlow logs, and tracks iteration times and
loss values for all the jobs in an app.
The allocation of a job changes over time and iteration times are
used to accurately estimate the placement sensitivity ($\mathcal{S}$)
from different GPU placements.  Loss values are used in our
Hyperband implementation to determine early stopping.
\name{}'s \arbiter{} is implemented as a separate module in YARN RM. 
We add gRPC-based interfaces between the \agent{} and the \arbiter{} 
to enable offers, bids, and final winning allocations. 
Further, the \arbiter{} tracks GPU leases to offer reclaimed GPUs as a part of the offers.

All the jobs we use in our evaluation are TensorFlow programs with
configurable hyper-parameters.  To
handle allocation changes at runtime, the
programs checkpoint model parameters to
HDFS every few iterations.  After a change in
allocation, they resume 
from the most recent checkpoint.

\section{Evaluation}
\label{sec:eval}

We evaluate \name{} on a $64$ GPU cluster and also use a event-driven simulator to model a larger $256$ GPU cluster. We compare against other state-of-the-art ML schedulers. 
Our evaluation shows the following key highlights -

\begin{figure}[t!]
	\subfloat[][Workload 1]{\includegraphics[width=0.49\columnwidth]{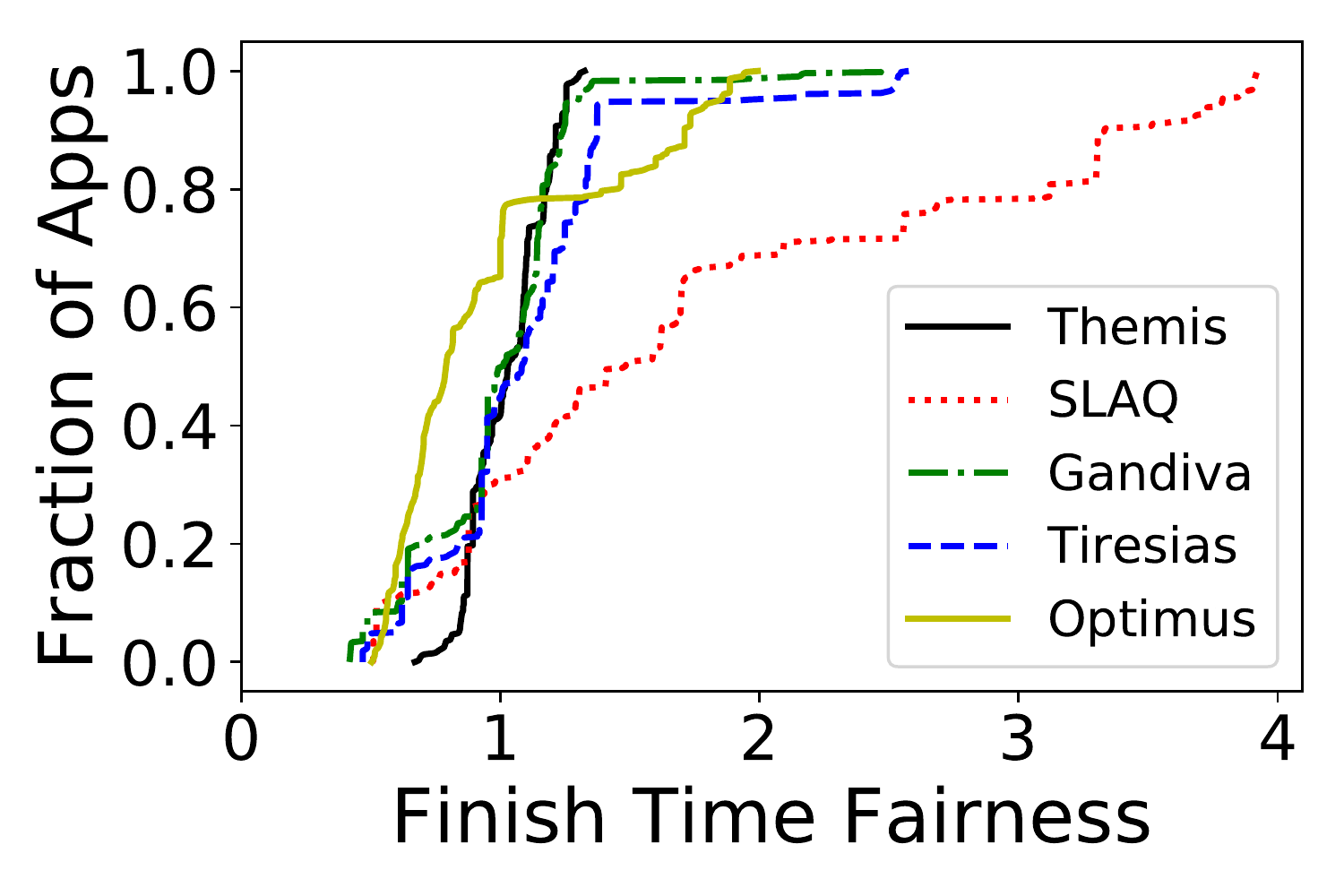}}
	\subfloat[][Workload 2]{\includegraphics[width=0.49\columnwidth]{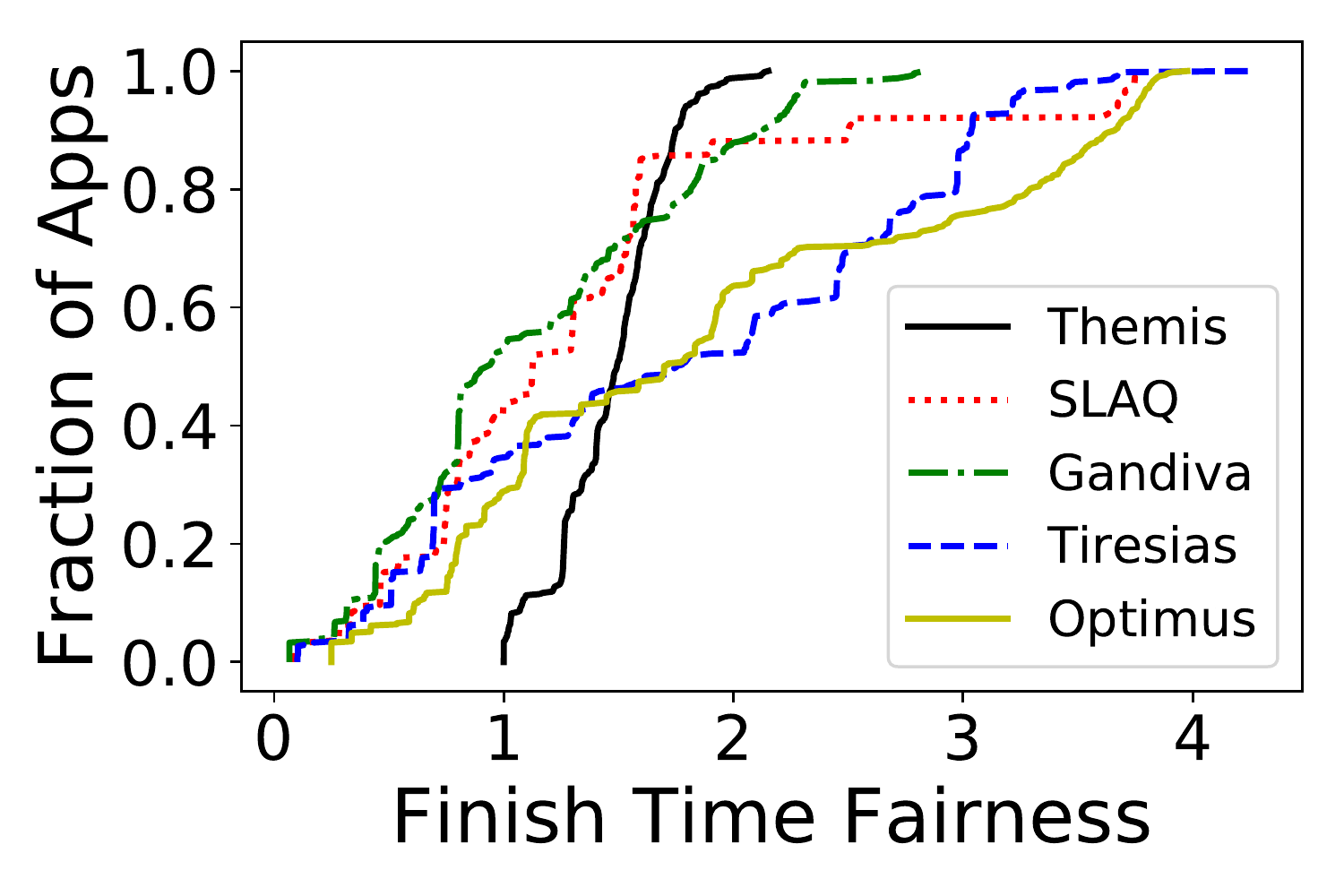}}
	\caption{\label{fig:ftf_comparison} [\textsc{Testbed}]  \footnotesize Comparison of finish-time
  fairness across schedulers}
\end{figure}

\begin{compactlist}
	\item  \name{} is better than other schemes on finish-time fairness while also offering better cluster efficiency
	(Figure~\ref{fig:ftf_comparison}-~\ref{fig:gpu_time_comparison}). 
	These experiments are done in our testbed.
	
	\item  \name{}'s benefits compared to other schemes improve with increasing fraction of placement sensitive apps and increasing contention in the cluster, and 
	these improvements hold even with errors -- random and strategic -- in finish-time fair metric estimations (Figure~\ref{fig:contention_effect} -~\ref{fig:truth-telling}).
	
	\item  \name{} enables a trade-off between finish-time fairness in the long-term and placement efficiency in the short-term.  
	Sensitivity analysis (Figure~\ref{fig:sensitivity_analysis}) 
	using simulations show that $f=0.8$ and a lease time of $10$ minutes gives maximum fairness while also utilizing the cluster efficiently. 
\end{compactlist}

\subsection{Experimental Setup}

\noindent{\bf Testbed Setup. } 
Our testbed is a $64$ GPU, $20$ machine cluster on Microsoft Azure~\cite{azure}. 
We use NC-series instances. We have $8$ NC12-series instances each with 
$2$ Tesla K80 GPUs and $12$ NC24-series instances each with $4$ Tesla K80 GPUs.

\noindent{\bf Simulator. } We develop an event-based simulator to evaluate
\name{} at large scale. 
The simulator assumes that estimates of the loss function curves for jobs are known ahead of time so as to predict the total number of iterations for the job.
Unless stated otherwise, all simulations are done on a 
heterogeneous $256$ GPU cluster. 
Our simulator assumes a $4$-level hierarchical locality model for GPU placements. 
Individual GPUs fit onto {\em slots} on {\em machines} occupying different cluster {\em racks}. 
 \footnote{The heterogeneous cluster consists of $16$ $8$-GPU machines $4$ slots and $2$ GPUs per slot), $6$ $4$-GPU machines ($4$ slots and $1$ GPU per slot), and $16$ $1$-GPU machines}

\noindent{\bf Workload. } 
We experiment with $2$ different traces that have different workload characteristics in both the simulator and 
the testbed - {\bf (i) Workload 1. } A publicly available trace of DNN training workloads at Microsoft~\cite{philly_analysis_atc,philly_traces}.  We scale-down the trace, using a two week snapshot and focusing on subset of jobs from the trace that correspond to hyper-parameter exploration jobs triggered by Hyperdrive.\label{hyperdrive_workload}
{\bf (ii) Workload 2. } We use the app arrival times from Workload 1,
generate jobs per app using the successive halving pattern characteristic of the Hyperband algorithm~\cite{hyperband},
and increase the number of tasks per job compared to Workload 1.
The distribution of number of tasks per job and number of jobs per app for the two workloads is shown in the Appendix~\ref{sec:appendix} (Figure~\ref{fig:eval_workload}).\label{hyperband_workload}

\noindent{\bf Baselines. } 
We compare \name{} against four state-of-the-art ML schedulers - Gandiva~\cite{gandiva}, Tiresias~\cite{tiresias}, Optimus~\cite{optimus}, and SLAQ~\cite{slaq}; these represent the best possible baselines for maximizing efficiency, maximizing fairness, minimizing average app completion time, and maximizing aggregate model quality, respectively. We implement these baselines in our testbed as well as the simulator as described below:

\noindent{\bf Ideal Efficiency Baseline - Gandiva.} Gandiva improves cluster utilization by packing jobs on as few machines as possible. 
In our implementation, Gandiva introspectively profiles ML job execution to infer placement preferences and migrates jobs to better meet these placement preferences. On any resource availability, all apps report their placement preferences and we allocate resources in a greedy highest preference first manner which has the effect of maximizing the average placement preference across apps. 
We do not model time-slicing and packing of GPUs as these system-level techniques can be integrated with \name{} as well and would benefit Gandiva and \name{} to equal extents. 

\noindent{\bf Ideal Fairness Baseline - Tiresias.} Tiresias defines a new service metric for ML jobs -- the aggregate GPU-time allocated to each job -- and allocates resources using the Least Attained Service (LAS) policy so that all jobs obtain equal service over time. In our implementation, on any resource availability, all apps report their service metric and we allocate the resource to apps that have the least GPU service. 
We do not implement the Gittin's index policy from Tiresias as it enables a trade-off between service-based fairness and average app completion time. Instead, we use a different baseline for average app completion time. 

\noindent{\bf Ideal Average App Completion Time Baseline - Optimus.} Optimus proposes the Shortest Remaining Time First (SRTF) policy for ML jobs. In our implementation, on any resource availability, all apps report their remaining time with allocations from the available resources and we allocate these resources using the SRTF policy.

\noindent{\bf Ideal Aggregate Model Quality - SLAQ.} SLAQ proposes a greedy scheme for improving aggregate model quality across all jobs. In our implementation, on any resource availability event, all apps report the decrease in loss value with allocations from the available resources and we allocate these resources in a greedy highest loss first manner.

\noindent{\bf Metrics. }  We use a variety of metrics to evaluate \name{}.

{\bf (i) Finish-time fairness: } We evaluate the fairness of
schemes by looking at the finish-time fair metric ($\rho$) distribution and the maximum value across apps. A tighter distribution and a lower value of maximum value of $\rho$ across apps indicate higher fairness. {\bf (ii) GPU Time: } We use \emph{GPU Time} as a measure of how efficiently the cluster is
utilized. 
For two scheduling schemes $S_1$ and $S_2$ that have GPU times $G_1$ and $G_2$ for executing the same amount of work, $S_1$ utilizes the cluster more efficiently than $S_2$ if $G_1 < G_2$.
{\bf (iii) Placement Score: } We give each allocation a placement score ($\le 1$). This is inversely proportional to slowdown, $\mathcal{S}$, that app experiences due to this allocation. The slowdown is dependent on the ML app properties and the network interconnects between the allocated GPUs. A placement score of $1.0$ is desirable for as many apps as possible.

\subsection{Macrobenchmarks}

In our testbed, we evaluate \name{} against all baselines on all the workloads.
We set the fairness knob value $f$ as $0.8$ and lease as $10$ minutes, which is 
informed by our sensitivity analysis results in Section~\ref{subsec:sensitivity_analysis}.

\begin{figure}[t!]
	\subfloat[][Workload 1]{\includegraphics[width=0.49\columnwidth]{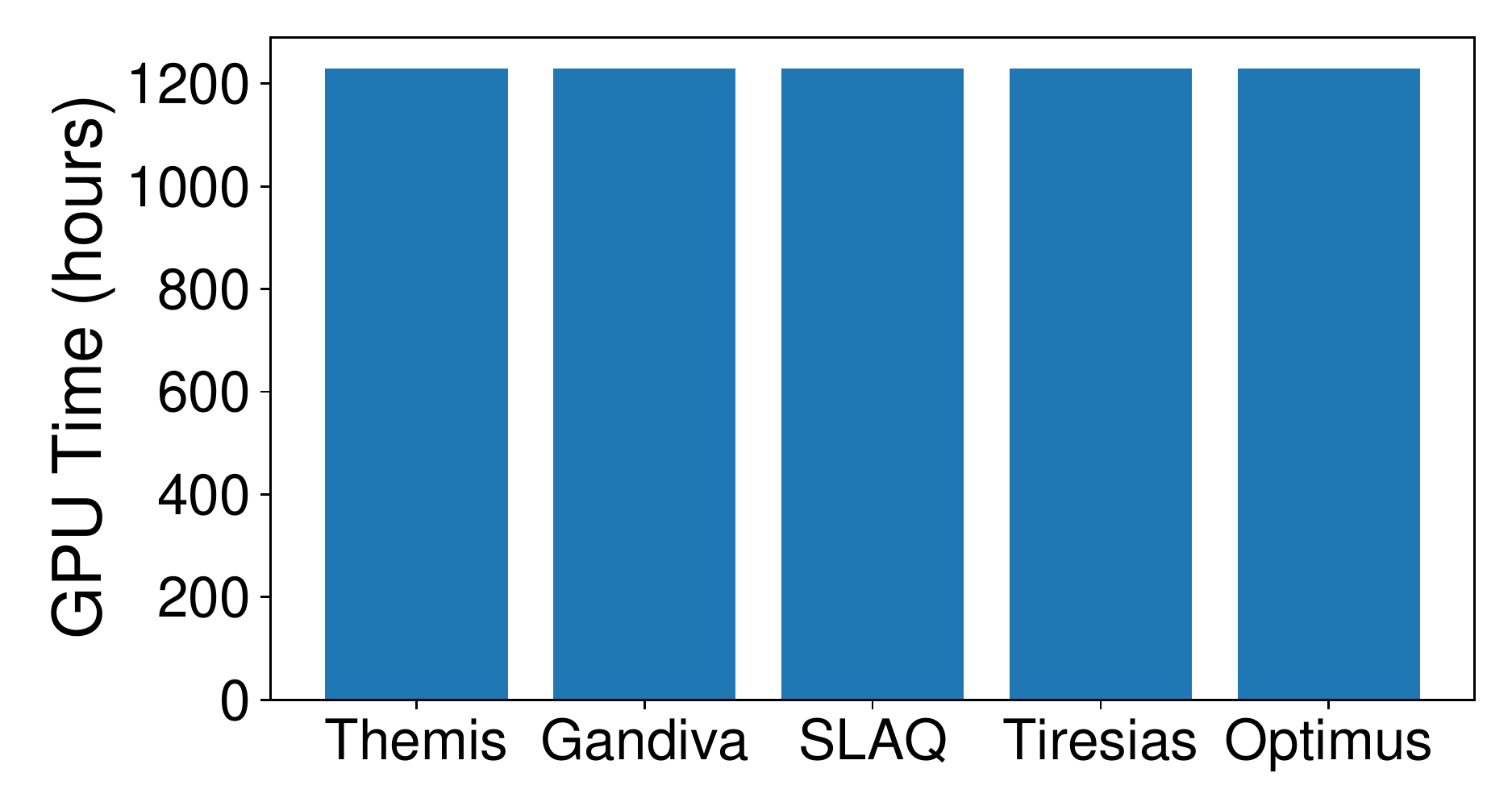}}
	\subfloat[][Workload 2]{\includegraphics[width=0.49\columnwidth]{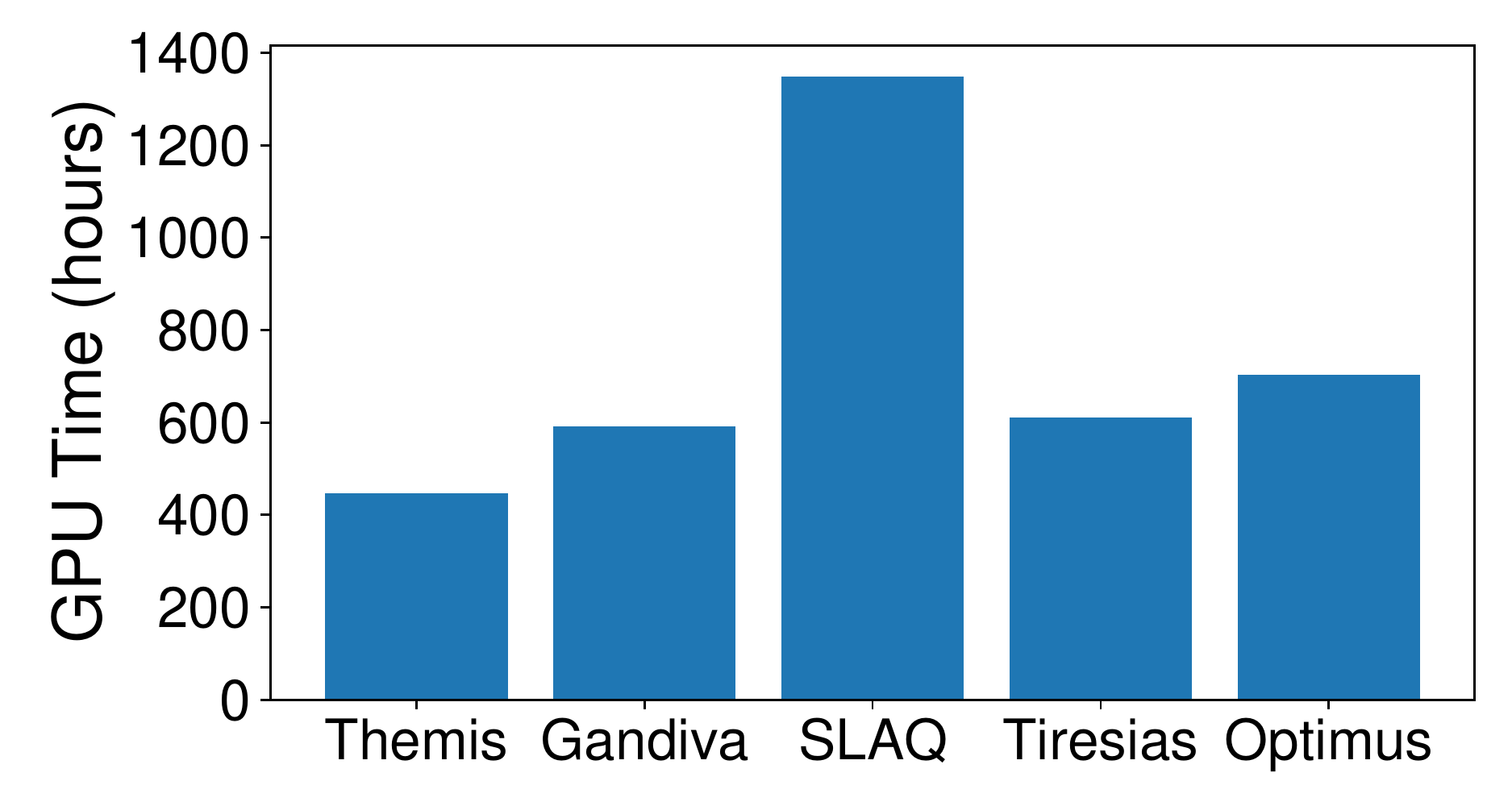}}
	\caption{\label{fig:gpu_time_comparison} [\textsc{Testbed}]  \footnotesize Comparison of total GPU times across schemes. Lower GPU time indicates better utilization of the GPU cluster}
\end{figure}

Figure~\ref{fig:ftf_comparison} shows the distribution of finish-time fairness metric, $\rho$, across apps for \name{} and all the baselines. 
We see that \name{} has a narrower distribution for the $\rho$ values which means that \name{} comes closest to giving all jobs an equal sharing incentive. Also, \name{} has the maximum $\rho$ value $2.2X$, $2.25X$, $1.75X$, and $3.25X$ better than Gandiva, Tiresias, Optimus, and SLAQ, respectively.
Figure~\ref{fig:gpu_time_comparison} shows a comparison of the efficiency in terms of the aggregate GPU time to execute the complete workload.
Workload 1 has similar efficiency across \name{} and the baselines as all jobs are either $1$ or $2$ GPU jobs and almost all allocations, irrespective of the scheme, end up as efficient. 
With workload 2, \name{} betters Gandiva by \textasciitilde 4.8\% and outperforms SLAQ by \textasciitilde 250\%.   
This improvement is because \name{} leads to global optimality in placement-driven packing due to the auction abstraction. 
Gandiva in contrast takes greedy locally optimal decisions.

\subsubsection{Sources of Improvement}\label{subsec:improvement_source}
In this section, we deep-dive into the reasons behind the wins in fairness and cluster efficiency in \name{}.

\begin{table}[t]
	\centering
	\begin{tabular}{ l | c | c | c | c}
		\hline
		\textbf{Job Type} & \textbf{GPU Time} & \textbf{\# GPUs} & \textbf{$\rho_{\name{}}$} & \textbf{$\rho_{Tiresias}$}\\
		\hline
		Long Job & \textasciitilde 580 mins & 4 & \textasciitilde 1 & \textasciitilde 0.9\\
		\hline
		Short Job  & \textasciitilde 83 mins & 2 & \textasciitilde 1.2 &
		\textasciitilde 1.9 \\
		\hline
	\end{tabular}
	\caption{[\textsc{Testbed}] Details of 2 jobs to understand the benefits of
  \name{}} %
	\label{tab:job_details}
\end{table}

Table~\ref{tab:job_details} compares the finish-time fair metric value
for a pair of short- and long-lived apps from our testbed run for
\name{} and Tiresias.  \name{} offers better sharing incentive for
both the short and long apps. \name{} induces altruistic behavior in
long apps. We attribute this to our choice of $\rho$ metric.  With
less than ideal allocations, even though long apps see an increase in
$T_{sh}$, their $\rho$ values do not increase drastically because of a
higher $T_{id}$ value in the denominator. Whereas, shorter apps see a
much more drastic degradation, and our round-by-round filtering of
farthest-from-finish-time fairness apps causes shorter apps to
participate in auctions more often.  Tiresias offers poor sharing
incentive for short apps as it treats short- and long-apps as the
same. This only worsens the sharing incentive for short apps.

\begin{figure}[!t]
	\begin{minipage}[b]{0.48\columnwidth}
		\includegraphics[width=\columnwidth]{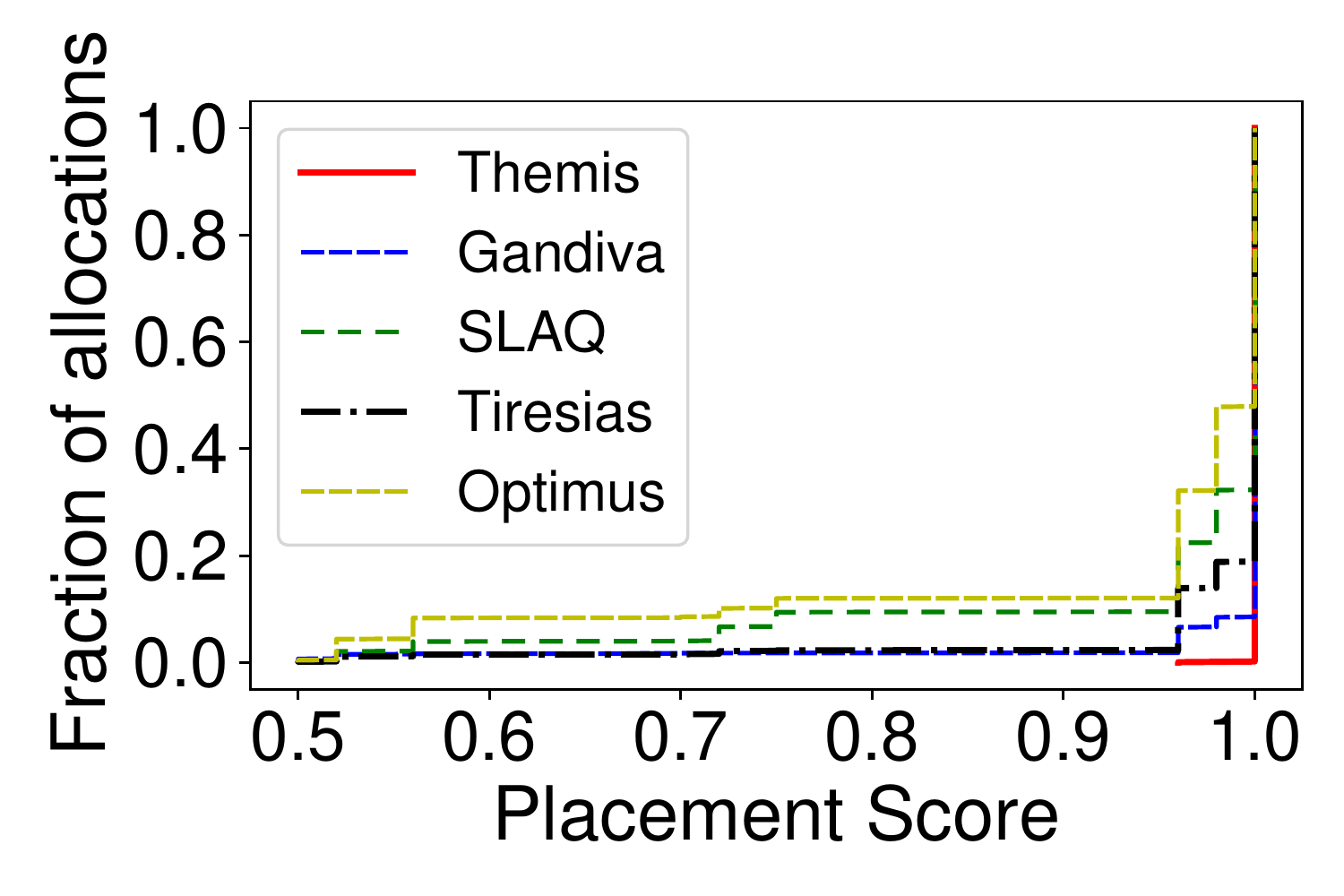}
		\caption{\textsc{Testbed}]  \footnotesize CDF of placement scores across schemes}
		\label{fig:placement_scores}
	\end{minipage}%
    \hspace{0.1cm}
    \begin{minipage}[b]{0.44\columnwidth}
    	\includegraphics[width=\columnwidth]{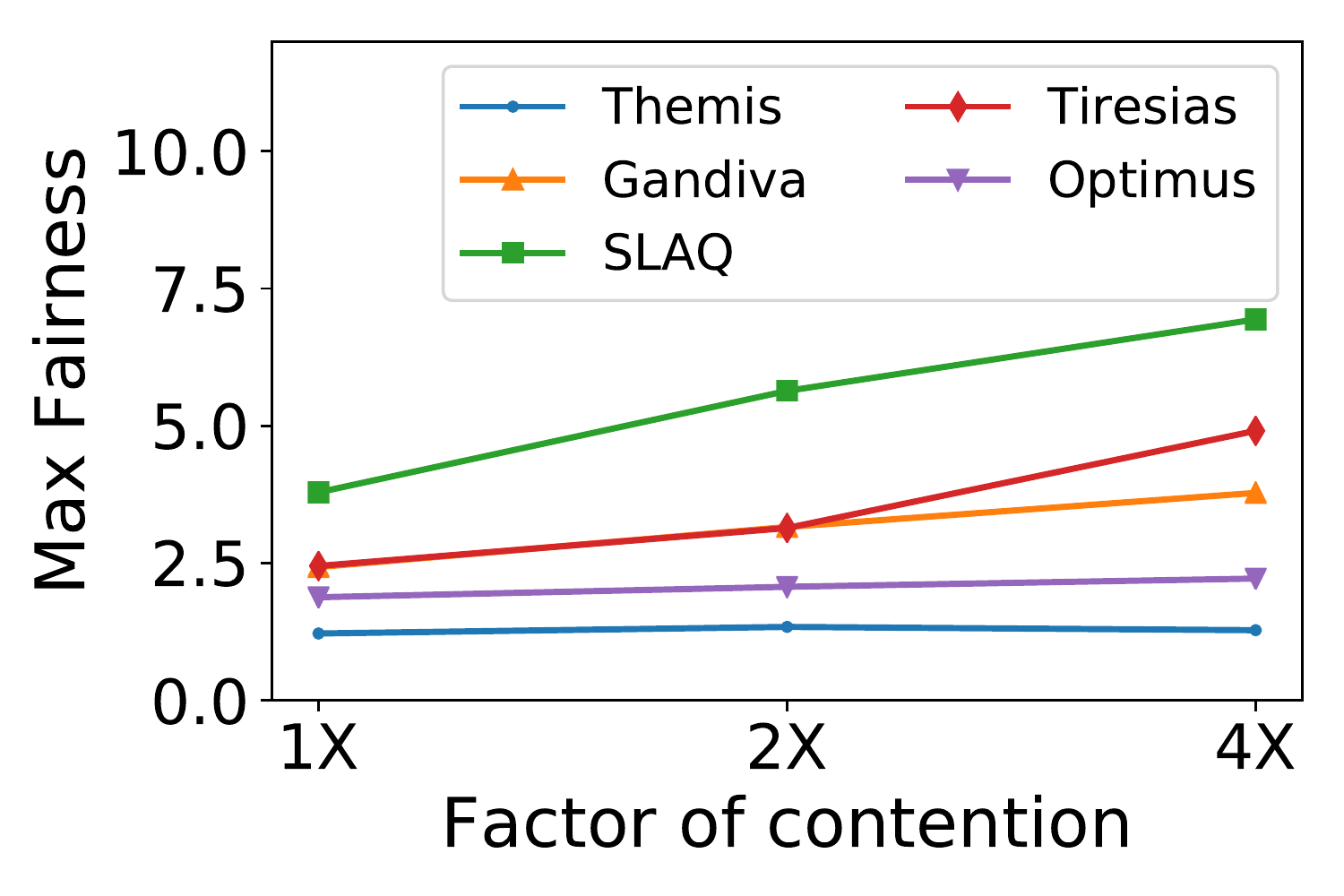}
    	\caption{[\textsc{Testbed}]  Impact of contention on finish-time fairness}
    	\label{fig:contention_effect}
    \end{minipage}
\end{figure}

Figure \ref{fig:placement_scores} shows the distribution of placement scores for all the schedulers. 
\name{} gives the best placement scores (closer to 1.0 is better) in workload 2, with Gandiva coming closest. 
Workload 1 has jobs with very low GPU demand and almost all allocations have a placement score of 1 irrespective of the scheme.
Tiresias, SLAQ, and Optimus are poor as they are agnostic of placement preferences. Gandiva performs
better but is not as efficient because of greedy local packing.

\subsubsection{Effect of Contention}

In this section, we analyze the effect of contention on finish-time fairness.  
We decrease the size of the cluster to half and quarter the original size to 
induce a contention of 2$X$ and 4$X$ respectively.
Figure \ref{fig:contention_effect} shows the change in max value of $\rho$ as the contention 
changes with workload 1. 
\name{} is the only scheme that maintains {\em sharing incentive} even in high contention scenarios. 
Optimus comes close as it preferably allocates resources to shorter apps. This behavior is similar to that in \name{}. \name{} induces altruistic shedding of resources by longer apps (Section~\ref{subsec:improvement_source}), giving shorter apps a preference in allocations during higher contention.

\subsubsection{Systems Overheads}

From our profiling of the experiments above, we find that each
\agent{} spends~29 (334) milliseconds to compute bids at the median
(95-\%). The 95 percentile is high because enumeration of possible
bids needs to traverse a larger search space when the number of
resources up for auction is high.

The \arbiter{} uses Gurobi~\cite{gurobi} to compute partial
allocation of resources to apps based on bids. This computation takes
~354 (1398) milliseconds at the median (95-\%ile). The high tail is once
again observed when both the number of offered resources and the
number of apps bidding are high. However, the time is small
relative to lease time.
The network overhead for communication between the \arbiter{} and individual apps
is negligible since we use the existing mechanisms used by
Apache YARN.

Upon receiving new resource allocations, the \agent{} changes
(adds/removes) the number of GPU containers available to its app. This change takes about ~$35$ ($50$)
seconds at the median ($95$-\%ile), i.e., 
 an overhead of
~$0.2$\% ($2$\%) of the app duration at the median ($95$-\%ile). 
Prior to relinquishing control over its
resources, each application must checkpoint its set of parameters. We
find that that this is model dependent but takes about 5-10 seconds on
an average and is driven largely by the overhead of check-pointing to
HDFS.

\subsection{Microbenchmarks}

\noindent{\bf Placement Sensitivity:}
We analyze the impact on finish-time fairness and cluster efficiency as the fraction of network-intensive apps in our workload increases. We synthetically construct 6 workloads and vary the percentage of network-intensive apps in these workloads across from $0$\%-$100$\%. 

\begin{figure}[t!]
	\centering
	\subfloat[Impact on Max $\rho$]{\includegraphics[width=0.49\columnwidth]{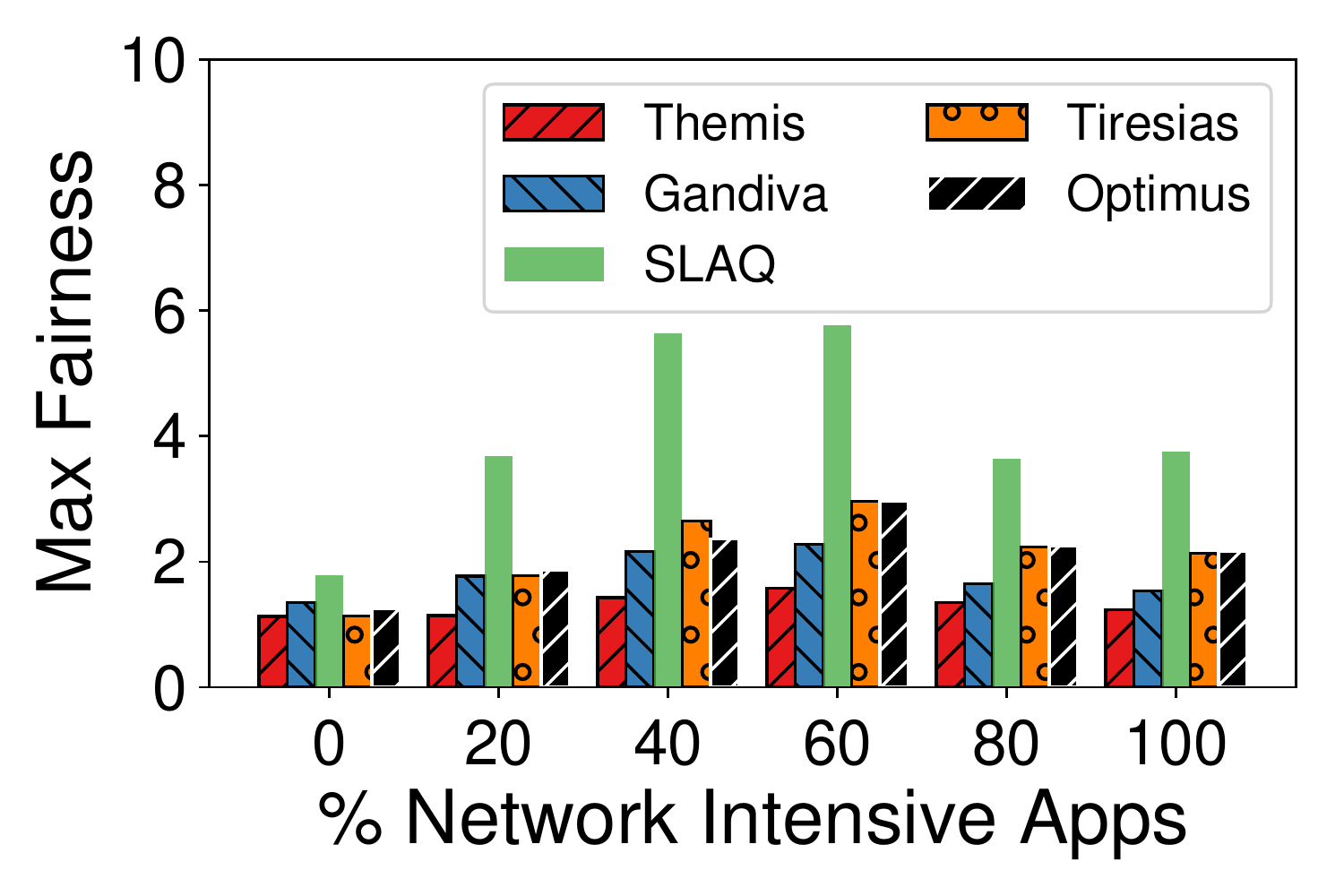}} \label{fig:placement_fairness}
	\subfloat[Impact on GPU Time]{\includegraphics[width=0.49\columnwidth]{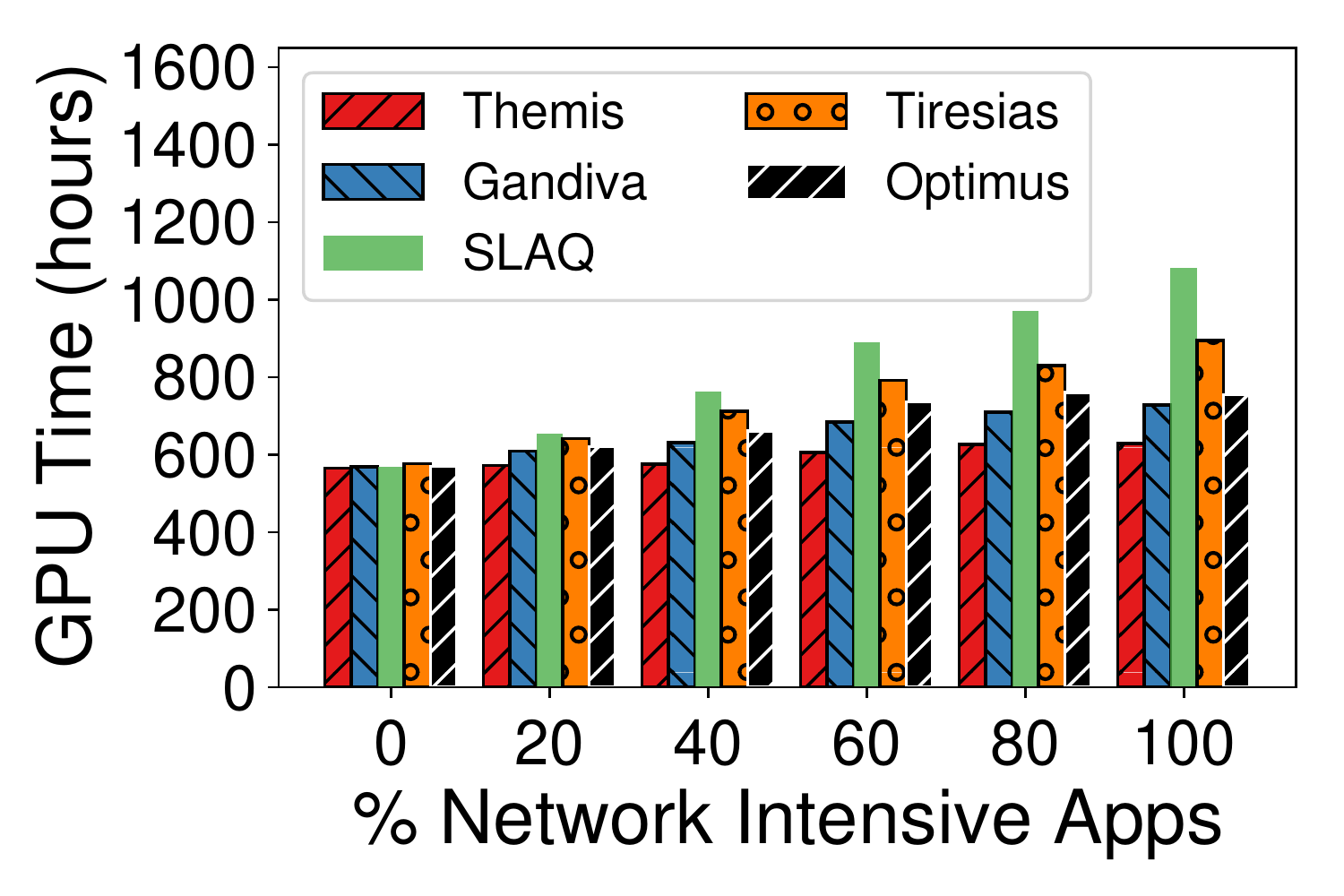}} \label{fig:placement_efficiency}
	\caption{\label{fig:placement_effect}
  [\textsc{Simulator}]
		Impact of placement sensitivity for varying compute-network apps distributions
	}
  \vspace{-0.05in}
\end{figure}

From Figure~\ref{fig:placement_effect}(a), we notice that {\em sharing incentive} degrades most when there is a heterogeneous mix of compute and network intensive apps (at 40\% and 60\%).
\name{} has a max $\rho$ value closest to $1$ across all these scenarios and is the only scheme to ensure this sharing incentive. 
When the workload consists solely of network-intensive apps, \name{} performs \textasciitilde
$1.24X$, $2.04X$, $1.72X$ and $1.77X$ better than Gandiva, SLAQ, Tiresias, and Optimus respectively
on max fairness. %

Figure \ref{fig:placement_effect}(b) captures the impact on cluster efficiency. 
With only compute-intensive apps, all scheduling schemes utilize the cluster equally efficiently. 
As the percentage of network intensive apps increases, \name{} has lower GPU times to execute the same workload. This means that \name{} utilizes the cluster more efficiently than other schemes.
In the workload with $100$\% network-intensive apps, \name{} performs \textasciitilde 8.1\% better than Gandiva (state-of-the-art for cluster efficiency).

\begin{figure}
	\begin{minipage}[t]{0.48\columnwidth}
		\includegraphics[width=\columnwidth]{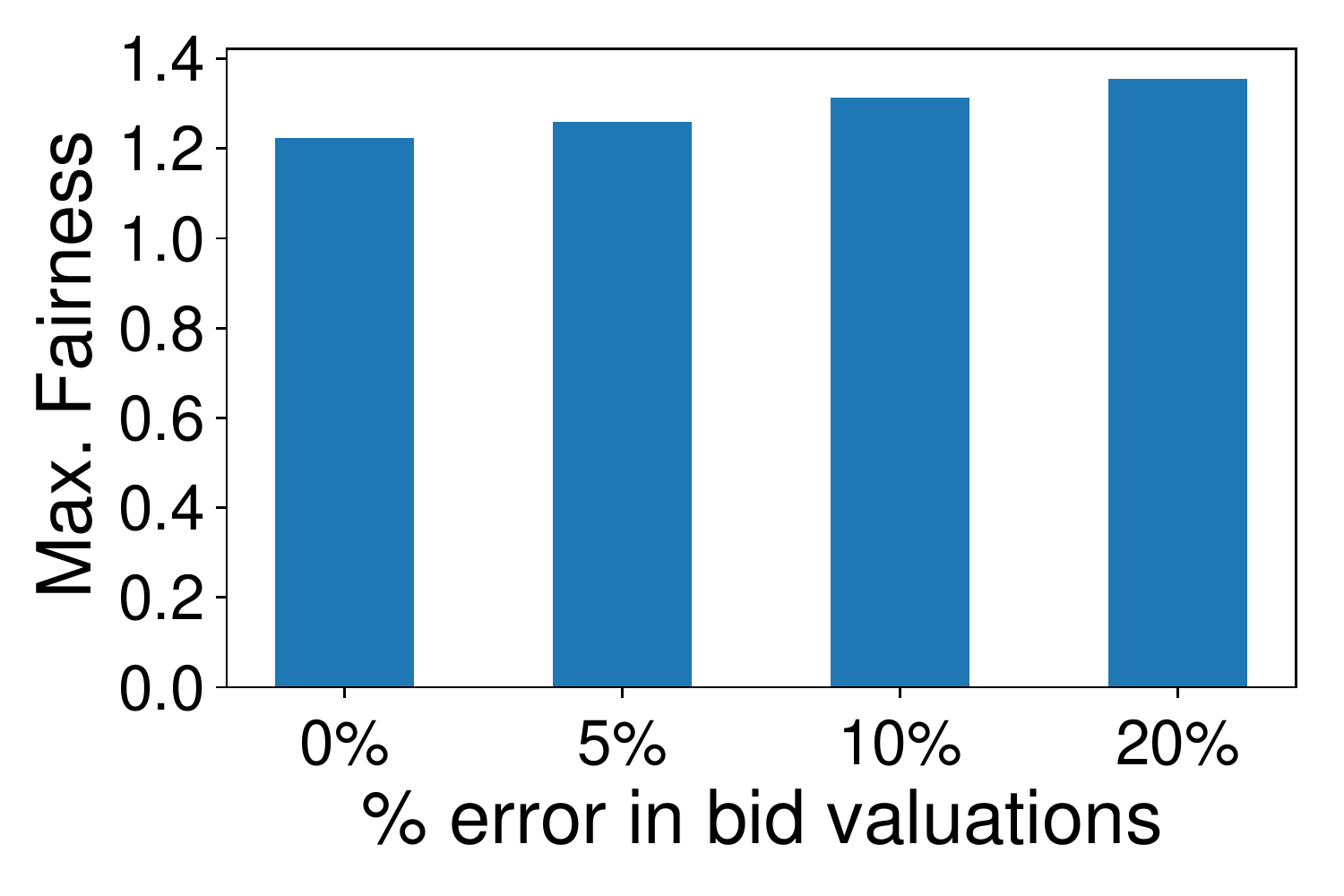}
		\caption{[\textsc{Simulator}]  Impact of error in bid values on max fairness}
		\label{fig:error_impact}
	\end{minipage}%
	\hspace{0.1cm}
	\begin{minipage}[t]{0.48\columnwidth}
		\includegraphics[width=\columnwidth]{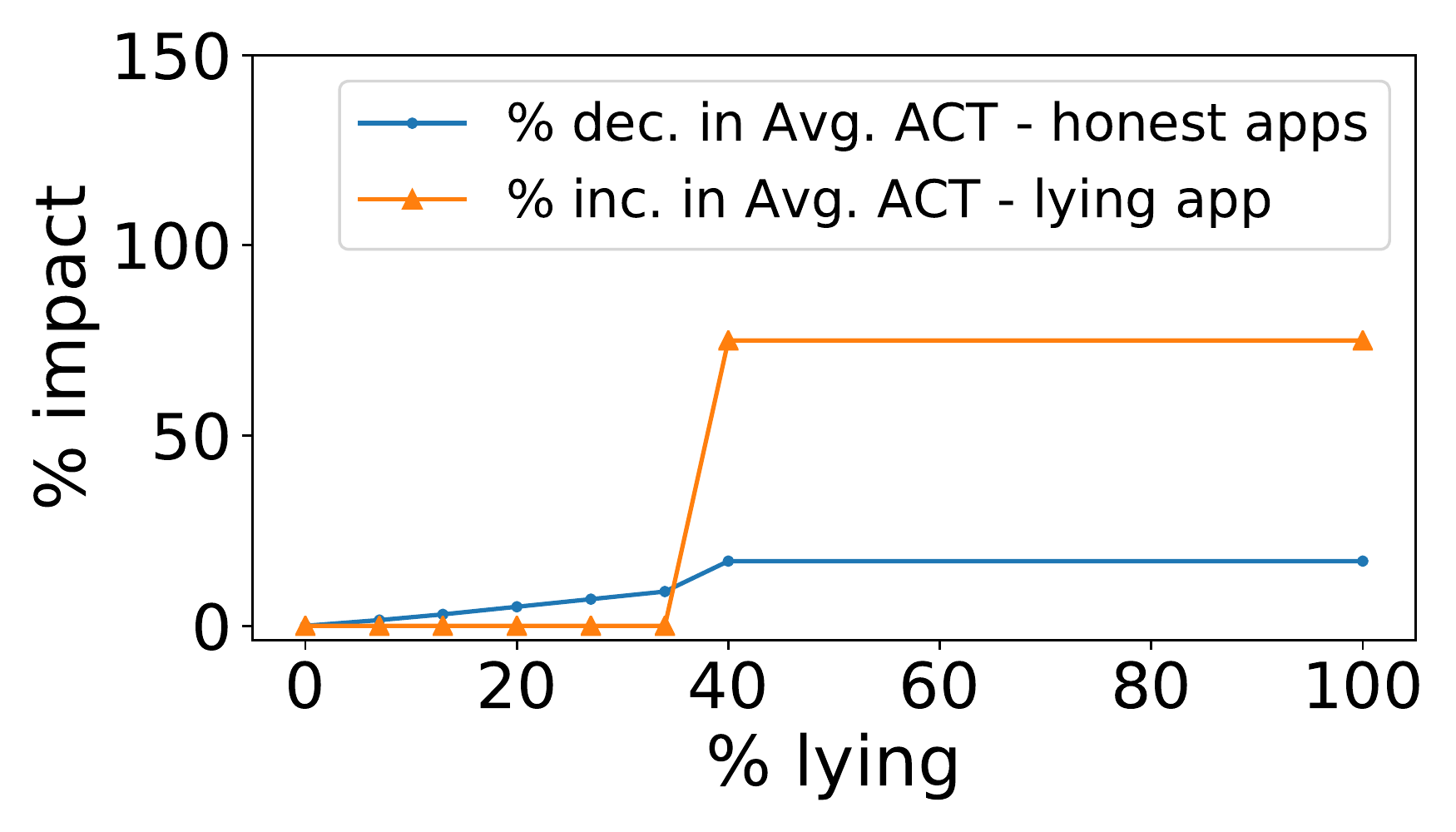}
		\caption{[\textsc{Simulator}] Strategic lying is detrimental}
		\label{fig:truth-telling}
	\end{minipage}	
\end{figure}

\noindent{\bf Error Analysis:} Here, we evaluate the ability of \name{} to handle errors in estimation of number of iterations and the slowdown ($\mathcal{S}$). For this experiment, we assume that all apps are equally susceptible to making errors in estimation. 
The percentage error is sampled at random from [-$X$, $X$] range for each app.
Figure \ref{fig:error_impact} shows the changes 
in max finish-time fairness as we vary $X$.
Even with $X = 20$\%, the change in max finish-time fairness is just $10.76$\% and is not significant.

\noindent{\bf Truth-Telling:}
To evaluate strategy-proofness, we use simulations. 
We use a cluster of $64$ GPUs with $8$ identical apps with equivalent placement preferences. 
The cluster has a single $8$ GPU machine and the others are all $2$ GPU machines. 
The most preferred allocation in this cluster is the $8$ GPU machine.
We assume that there is a single strategically lying app and $7$ truthful apps. 
In every round of auction it participates in, the lying app over-reports the slowdown 
with staggered machine placement or under-reports the slowdown with dense machine placement 
by $X$\%. Such a strategy would ensure higher likelihood of winning the $8$ GPU machine. 
We vary the value of $X$ in the range $[0, 100]$ and analyze the lying app's 
completion time and the average app completion time of the truthful apps in 
Figure~\ref{fig:truth-telling}. 
We see that at first the lying app does not experience any decrease in its own 
app completion time. On the other hand, we see that the truthful apps do better 
on their average app completion time. This is because the hidden payment from the partial 
allocation mechanism in each round of the auction for the lying app remains the same 
while the payment from the rest of the apps keeps decreasing. 
We also observe that there is a sudden tipping point at $X > 34$\%. 
At this point, there is a sudden increase in the hidden payment for 
the lying app and it loses a big chunk of resources to other apps. 
In essence, \name{} incentivizes truth-telling.

\subsection{Sensitivity Analysis}
\label{subsec:sensitivity_analysis}
\begin{figure}[t!]
	\centering
	\subfloat[Impact on Max Fairness]{\includegraphics[width=0.49\columnwidth]{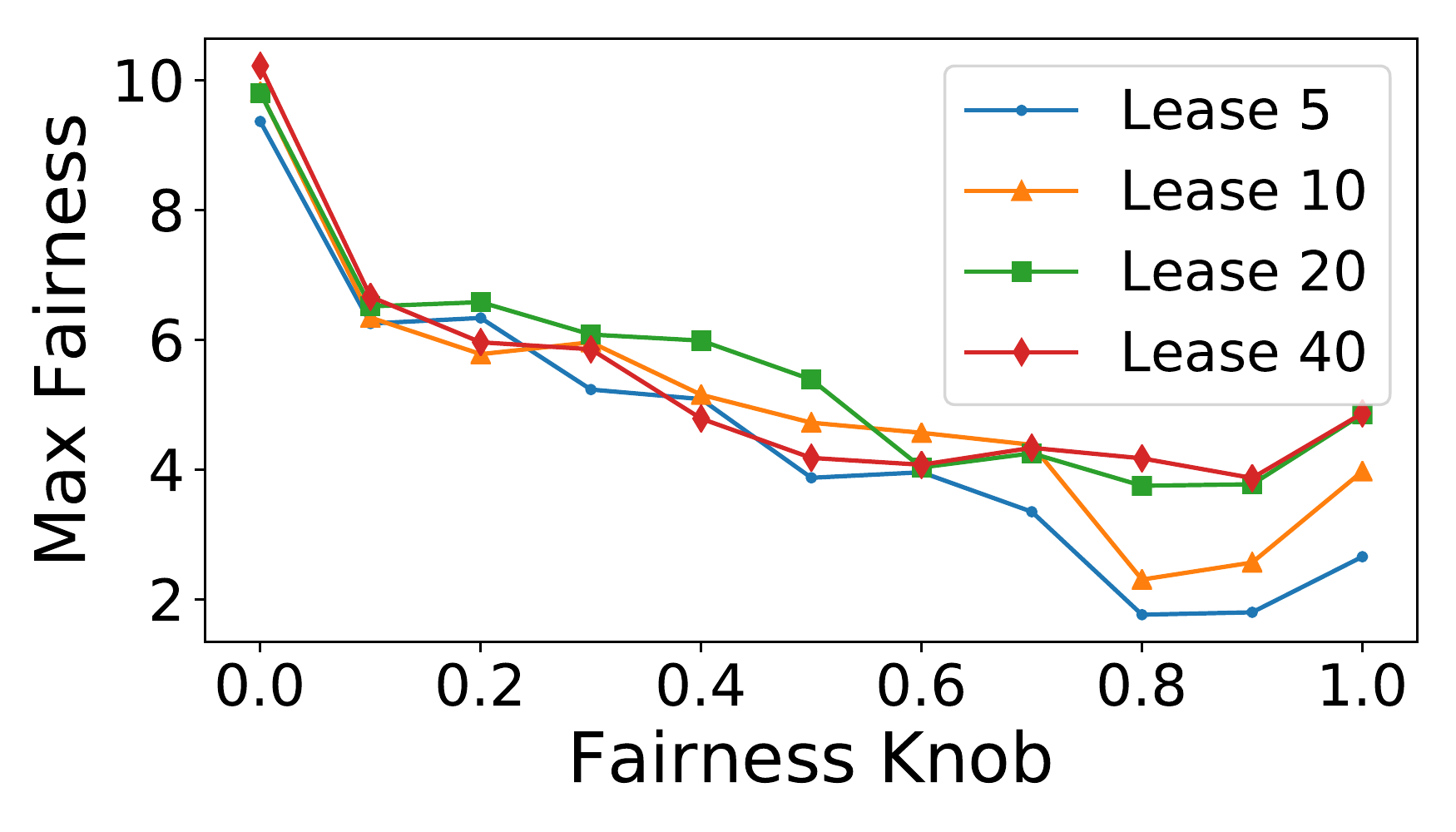}} \label{fig:sens_fairness}
	\subfloat[Impact on GPU Time]{\includegraphics[width=0.49\columnwidth]{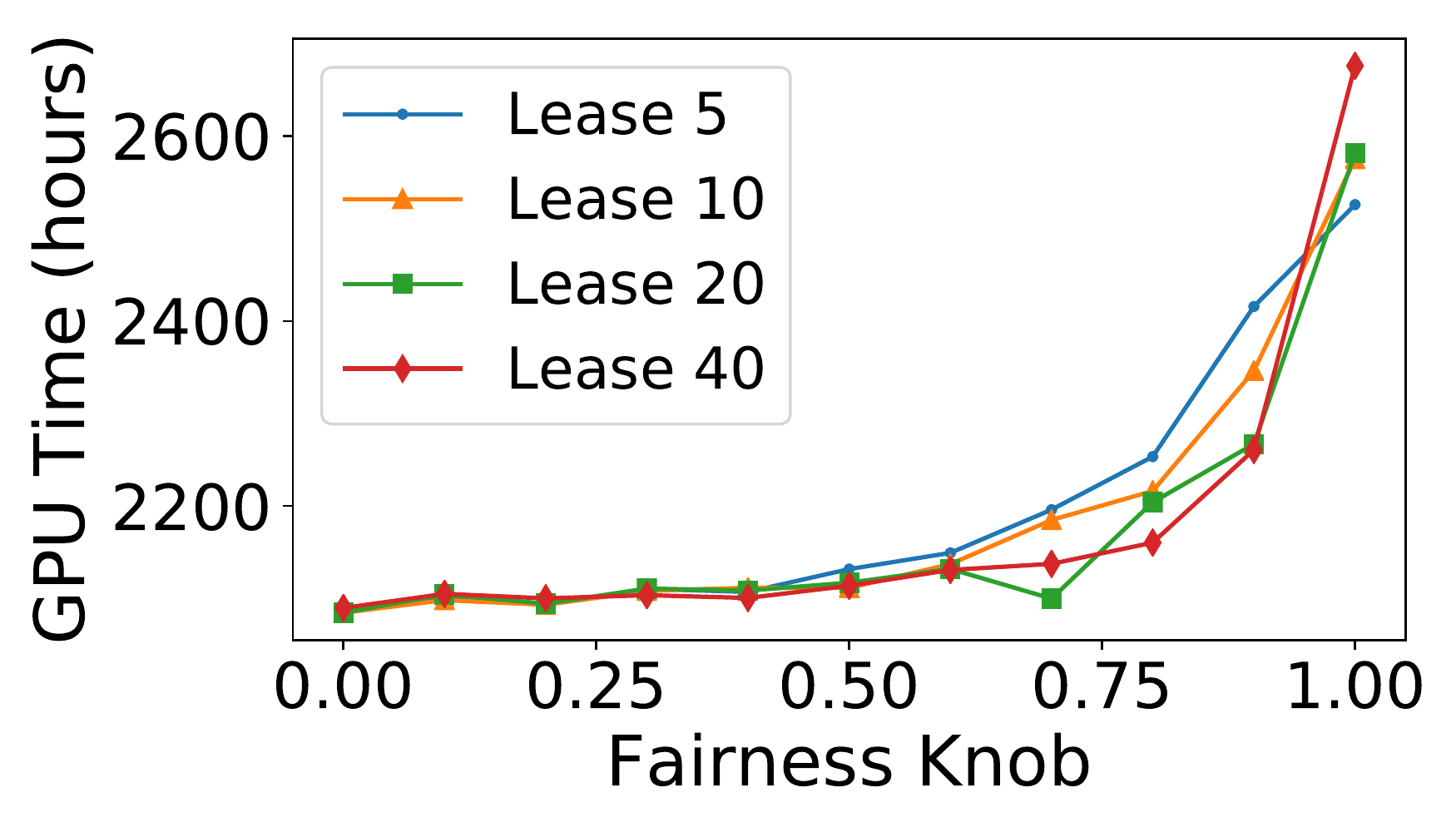}} \label{fig:sens_efficiency}
	\caption{\label{fig:sensitivity_analysis} [\textsc{Simulator}] 
		\footnotesize
    Sensitivity of fairness knob and lease time.
	}   
\end{figure}

We use simulations to study \name{}'s sensitivity to fairness knob $f$ and
the lease time.
Figure~\ref{fig:sensitivity_analysis} (a) shows the impact on max $\rho$ as we vary the fairness knob $f$. We observe that filtering $(1-f)$ fraction of apps helps with ensuring better {\em sharing incentive}. As $f$ increases from $0$ to $0.8$, we observe that fairness improves. Beyond $f = 0.8$, max fairness worsens by around a factor of $1.5X$. 
We see that the quality of sharing incentive, captured by max $\rho$, degrades at $f=1$ because we observe that only a single app with highest $\rho$ value participates in the auction. This app is forced sub-optimal allocations because of poor placement of available resources with respect to the already allocated resources in this app.
We also observe that smaller lease times promote better fairness since frequently filtering apps reduces the time that queued apps wait for an allocation.

Figure \ref{fig:sensitivity_analysis} (b) shows the impact on the efficiency of cluster usage as we
vary the fairness knob $f$. We observe that the efficiency decreases as the value of $f$ increases.
This is because the number of apps that can bid for an offer reduces as we increase $f$
leading to fewer opportunities for the \arbiter{} to pack jobs efficiently. Lower lease values mean
than models need to be check-pointed more often (GPUs are released on lease expiry) and hence higher
lease values are more efficient.

Thus we choose $f=0.8$ and $lease=10$ minutes.

\section{Related Work}

Cluster scheduling for ML workloads has been targeted by a number of recent works including
SLAQ~\cite{slaq}, Gandiva~\cite{gandiva}, Tiresias~\cite{tiresias} and Optimus~\cite{optimus}. These
systems target different objectives and we compare against them in Section~\ref{sec:eval}.

We build on rich literature on cluster scheduling
disciplines~\cite{drf,graphene,carbyne,tetris} and two level
schedulers~\cite{mesos,borg,omega}. While those disciplines/schedulers
don't apply to our problem, we build upon some of their ideas, e.g.,
 resource offers in~\cite{mesos}. 
Sharing incentive was outlined by DRF~\cite{drf}, 
but we focus on long term fairness with our finish-time  metric.
Tetris~\cite{tetris} proposes resource-aware packing with an option to trade-off for fairness using multi-dimensional bin-packing as the mechanism for achieving that. In
\name{}, we instead focus on fairness with an option to trade-off for 
placement-aware packing, and use auctions as our mechanism.

Some earlier schemes~\cite{carbyne,graphene} also attempted to emulate
the long term effects of fair allocation. Around occasional barriers,
unused resources are re-allocated across jobs.  \name{} differs in
many respects: First, earlier systems focus on batch
analytics. Second, earlier schemes rely on instantaneous
resource-fairness (akin to DRF), which has issues with
placement-insensitivity and not accounting for long tasks. Third, in
the ML context there are no occasional barriers. While barriers do arise
due to synchronization of parameters in ML jobs, they happen at {\em
  every} iteration. Also, resources unilaterally given up by a job may
not be usable by another job due to placement sensitivity.

\section{Conclusion}

In this paper we presented \name{}, a fair scheduling framework for ML training workloads. We showed
how existing fair allocation schemes are insufficient to handle long-running tasks and placement
preferences of ML workloads. To address these challenges we proposed a new long term fairness
objective in finish-time fairness. We then presented a two-level semi-optimistic scheduling
architecture where ML apps can bid on resources offered in an auction. Our experiments show that 
\name{} can improve fairness {\em and} efficiency compared to state of the art
schedulers.

\bibliographystyle{abbrv}
\bibliography{themis}

\appendix
\section{Appendix}
\label{sec:appendix}

\begin{proofof}{Theorem 3.1}
    Examples in Figure~\ref{fig:drf-violates} and Section~\ref{sec:placement_pref} shows that DRF violates SI, EF, and PE. 
    Same examples hold true for LAS policy in Tiresias. The service metric \ie~the GPU in Instance 1 and Instance 2 is the same for A1 and A2 in terms of LAS and is deemed a fair allocation over time. However, Instance 1 violates SI as A1 (VGG16) and A2 (VGG19) would prefer there own independent GPUs and Instance 2 violates EF and PE as A2 (VGG19) prefers the allocation of A1 (Inceptionv3) and PE as the optimal allocation after taking into account placement preferences would interchange the allocation of A1 and A2.
\end{proofof}

\begin{proofof}{Theorem 3.2}
    We first show that the valuation function, $\rho(.)$, for the case of ML jobs is homogeneous. 
    This means that $\rho(.)$ has the following property: $\rho(m*\myvec{G})$ = $m*\rho{\myvec{G}}$. 
    
    Consider a job with GPUs spread across a set of some $M$ machines. If we keep this set of machines the same, and increase the number of GPUs allocated on these same set of machines by a certain factor then the shared running time ($T_{sh}$) of this job decreases proportionally by the same factor. This is so because the slowdown, $\mathcal{S}$ remains the same. Slowdown is determined by the slowest network interconnect between the machines. The increased allocation does not change the set of machines $M$. The independent running time ($T_{id}$) remains the same. This means that $\rho$ also proportionally changes by the same factor. 
    
    Given, homogeneous valuation functions, the PA mechanism guarantees SP, PE and EF~\cite{nopaymentfair}. However, PA violates SI due to the presence of hidden payments. This also make PA not work-conserving.  
\end{proofof}

\begin{proofof}{Theorem 3.3}
    With multi-round auctions we ensure truth-telling of $\rho$ estimates in the visibility phase. This is done by the \agent by using the cached $\rho(.)$ estimates from the last auction the app participated in. In case an app gets leftover allocations from the leftover allocation mechanism, the \agent updates the $\rho$ estimate again by using the cached $\rho(.)$ table. In this way we guarantee SP with multi-round auctions. 
    
    As we saw in Theorem 3.2, an auction ensures PE and EF. In each round, we allocate all available resources using auctions. This ensures end-to-end PE and EF. 

    For maximizing sharing incentive, we always take a fraction $1-f$ of apps in each round. A wise choice of $f$ ensures that we filter in all the apps with $\rho > 1$ that have poor sharing incentive. We only auction the resources to such apps which maximizes sharing incentive.
\end{proofof}

\subsection{Workload Details}
\begin{figure}[t!]
	\subfloat[][CDF GPUs per job]{\includegraphics[width=0.49\columnwidth]{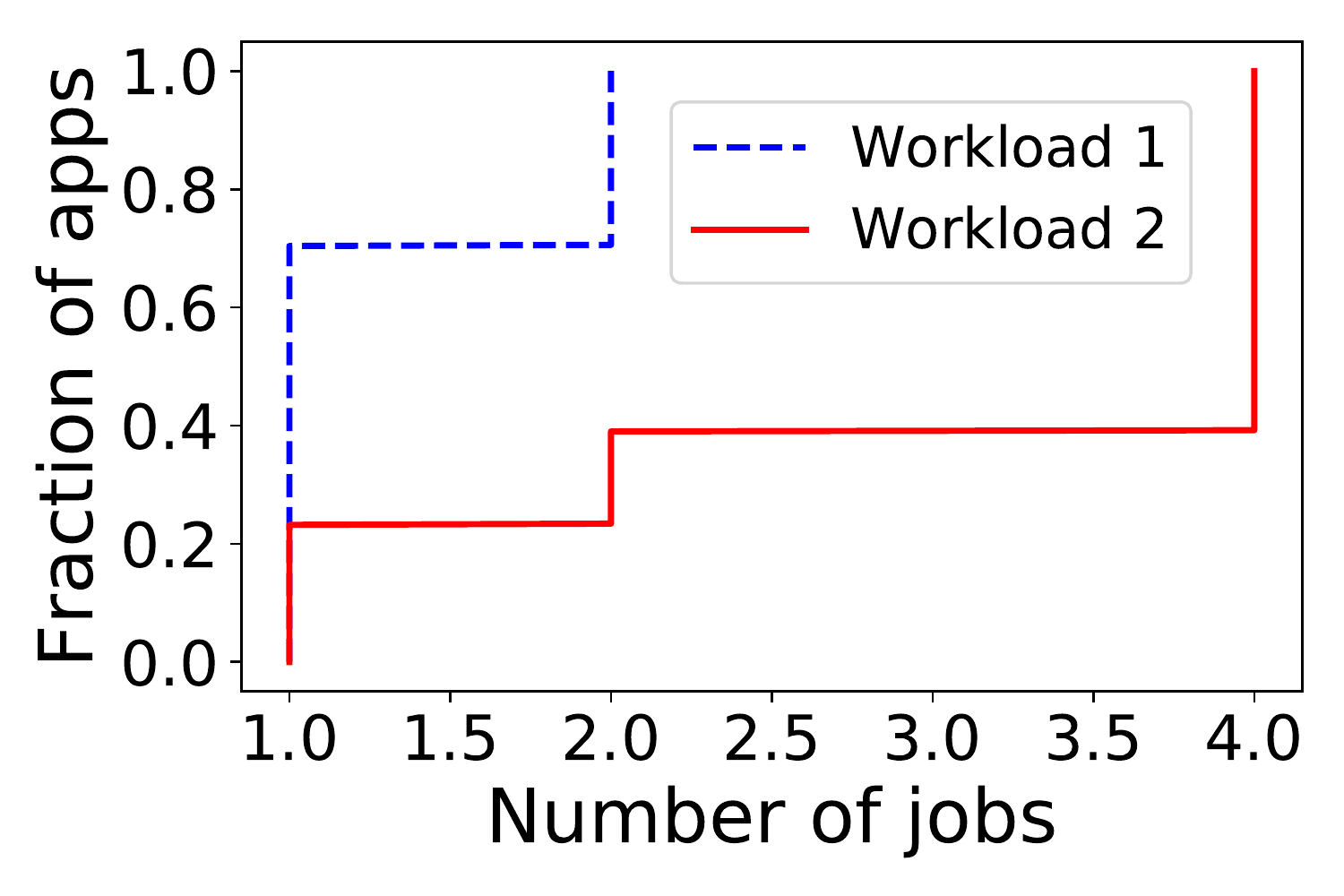}}
	\subfloat[][CDF jobs per app]{\includegraphics[width=0.49\columnwidth]{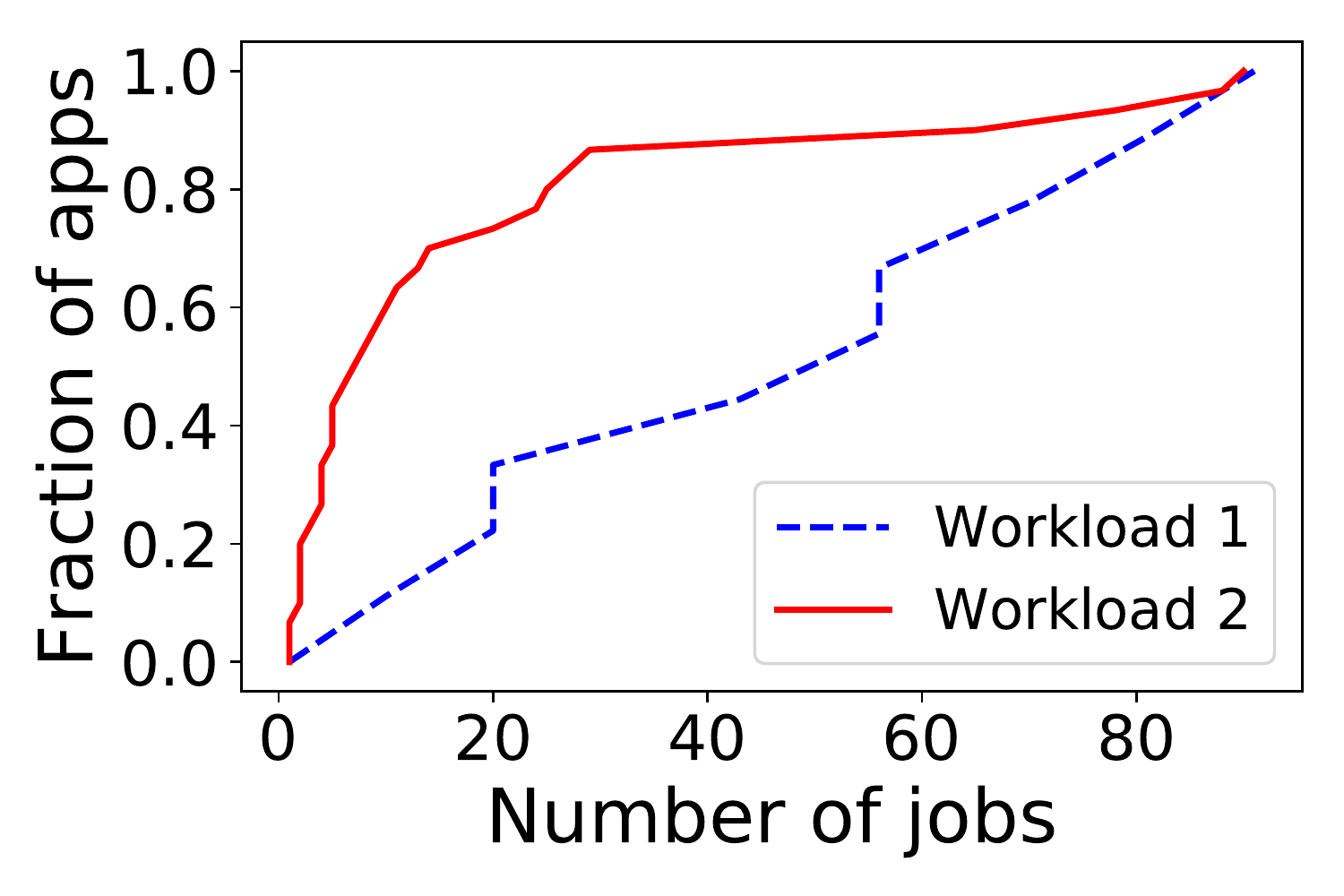}}
	\caption{\label{fig:eval_workload} \footnotesize Details of 2 workloads used for evaluation of \name{}}
\end{figure}

We experiment with $2$ different traces that have different workload characteristics in both the simulator and 
the testbed - {\bf (i) Workload 1. } A publicly available trace of DNN training workloads at Microsoft~\cite{philly_analysis_atc,philly_traces}.  We scale-down the trace, using a two week snapshot and focusing on subset of jobs from the trace that correspond to hyper-parameter exploration jobs triggered by Hyperdrive.\label{hyperdrive_workload}
{\bf (ii) Workload 2. } We use the app arrival times from Workload 1,
generate jobs per app using the successive halving pattern characteristic of the Hyperband algorithm~\cite{hyperband}.
The distribution of number of tasks per job and number of jobs per app for the two workloads is shown in Figure~\ref{fig:eval_workload}.

The traces comprise of models from three categories - computer vision (CV - 10\%), natural language processing (NLP - 60\%) and speech (Speech - 30\%). We use the same mix of models for each category as outlined in Gandiva~\cite{gandiva}. We summarize the models in Table~\ref{model-mixture}. 

\begin{center}
\begin{table}
    \begin{tabular}{ |c|c|c|c| } 
    \hline
     & Model & Type & Dataset \\
    \hline
    \multirow{5}{1.2em}{10\%} & Inceptionv3~\cite{inceptionv3} & CV & ImageNet~\cite{imagenet} \\ 
    & Alexnet~\cite{alexnet} & CV & ImageNet \\ 
    & Resnet50~\cite{resnet50} & CV & ImageNet \\
    & VGG16~\cite{vgg} & CV & ImageNet \\
    & VGG19~\cite{vgg} & CV & ImageNet \\ 
    \hline
    \multirow{4}{1.2em}{60\%} & Bi-Att-Flow~\cite{bi-att-flow} & NLP & SQuAD~\cite{squad} \\ 
    & LanguageModel~\cite{languagemodel} & NLP & PTB~\cite{ptb} \\ 
    & GNMT~\cite{gnmt} & NLP & WMT16~\cite{wmt16} \\
    & Transformer~\cite{transformer} & NLP & WMT16 \\
    \hline
    \multirow{2}{1.2em}{30\%} & Wavenet~\cite{wavenet} & Speech & VCTK~\cite{vctk} \\ 
    & DeepSpeech~\cite{deepspeech} & Speech & CommonVoice~\cite{commonvoice} \\ 
    \hline
    \end{tabular}
    \caption{\label{model-mixture} \footnotesize Models used in our trace.}
\end{table}
\end{center}

\end{document}